\documentstyle[aps,prb,epsfig,float,twocolumn]{revtex}

\newcommand{\be}{\begin{equation}}
\newcommand{\ee}{\end{equation}}
\newcommand{\bea}{\begin{eqnarray}}
\newcommand{\eea}{\end{eqnarray}}
\newcommand{\rd}{\mbox{d}}
\newcommand{\p}{\partial}

\begin{document}
\twocolumn[\hsize\textwidth\columnwidth\hsize\csname
@twocolumnfalse\endcsname

\title{Quantum theory of bilayer quantum Hall smectics}

\author{Emiliano Papa$^a$, John Schliemann$^{a,b}$, A.~H. MacDonald$^a$,
and Matthew P.~A. Fisher$^c$}

\address{$^a$Department of Physics, The University of Texas, Austin, TX 78712}
\address{$^b$Department of Physics and Astronomy, University of Basel, 
CH-4056 Basel, Switzerland}
\address{$^c$Kavli Institute for Theoretical Physics, University of California
at Santa Barbara, Santa Barbara CA 93106}

\date{\today}

\maketitle

\begin{abstract}

Mean-field theory predicts that bilayer quantum Hall systems at odd integer total filling factors 
can have stripe ground states in which the top Landau level is occupied alternately by electrons 
in one of the two layers.  We report on an analysis of the properties of these states 
based on a coupled Luttinger liquid description that is able to account for 
quantum fluctuations of charge-density and position along each stripe edge.  
The soft modes associated with the broken symmetries of the stripe state lead 
to an unusual coupled Luttinger liquid system with strongly enhanced low-temperature heat
capacity and strongly suppressed low-energy tunneling density of states.   
We assess the importance of the 
intralayer and interlayer back-scattering terms in the microscopic Hamiltonian, 
which are absent in the Luttinger liquid description, by employing a 
perturbative renormalization group approach which re-scales
time and length along but not transverse to the stripes. 
With interlayer back-scattering interactions present
the Luttinger liquid states are unstable either to an incompressible 
striped state that has spontaneous interlayer phase coherence
and a sizable charge gap even at relatively large layer separations, or
to Wigner crystal states.  Our quantitative estimates of the gaps produced
by back-scattering interactions are summarized in Fig.~\ref{phase_diagram} by a schematic
phase diagram intended to represent predicted experimental findings in very high mobility 
bilayer systems at dilution refrigerator temperatures as a function of layer separation and 
bilayer density balance.
We predict that the bilayer will form incompressible isotropic interlayer phase coherent states
for small layer separations, say $d \le 1.5 \ell$.
At larger interlayer spacings, however, the bilayer will tend to form one of
several different anisotropic states depending on the layer charge balance,
which we parameterize by the fractional filling factor $\nu$ contributed by one of the two layers. 
For large charge imbalances ($\nu$ far from $1/2$), we predict states in which 
anisotropic Wigner crystals form in each of the layers.  For $\nu$ closer to $1/2$, we 
predict stripe states that have spontaneous inter-layer phase coherence and a 
gap for charged excitations.  These states should exhibit the quantum Hall effect
for current flowing within the layers and also the 
giant interlayer tunneling conductance anomalies 
at low bias voltages that have been observed in bilayers when the $N=0$ Landau level
is partially filled. 
When the gaps produced by backscattering interactions are sufficiently small, 
the phenomenology observed at typical dilution fridge temperatures 
will be that of a smectic metal, anisotropic transport without a quantum Hall effect.
For stripe states in the $N =2$ Landau level, this behavior is expected 
over a range of bilayer charge imbalances on both sides of $\nu=1/2$.

\end{abstract}
\vskip2pc]


\section{Introduction}

The recent discovery of strongly anisotropic transport 
in single layer quantum Hall systems near half-odd
integer filling factors \cite{Lilly,Du,Shayegan} has attracted much
experimental \cite{Eisenstein} and theoretical \cite{Fogler} interest. 
Transport anisotropies have been observed in single two-dimensional (2D) electron
gas layers at half filling of 
Landau levels with index $N\geq 2$, {\it i.e.} at filling factors
$\nu=9/2, 11/2, \dots$. This effect is commonly ascribed to the formation 
of striped charge-density-wave phases, predicted on the basis of Hartree-Fock 
calculations by Koulakov {\em et al.} 
\cite{Koulakov} and by Moessner and Chalker \cite{Moessner} with additional 
theoretical support from 
subsequent exact-diagonalization \cite{Rezayi} and DMRG \cite{Yoshioka} 
numerical studies.  The stripe state is a consequence of the form factors that arise in
describing interactions between electrons in higher kinetic energy 
Landau level orbitals and allow density waves to form in cyclotron-orbit-center
coordinates that have a very small electron-density-wave amplitude and 
therefore little electrostatic energy penalty.   

The physics of quantum Hall systems is enriched by the additional
degrees of freedom that appear in bilayer systems\cite{Pinczuk}
in which two 2D electron layers have a separation $d$ small enough that their
interactions have consequences.  For total filling factor $\nu_{T} = 1$ and other odd
integer total filling factors, 
interlayer interactions can lead to a state with spontaneous phase coherence \cite{spontcoh} 
between the layers and a charge gap that is revealed experimentally \cite{Murphy}
by the quantum Hall effect.  Further spectacular experimental manifestations 
of spontaneous phase coherence were 
revealed very recently in 2D to 2D tunneling and Hall drag experiments by 
Eisenstein and collaborators. \cite{jpecoherence}.  In tunneling studies spontaneous
coherence is signaled by a sharp 
zero bias peak in the differential conductance between the layers.
As the ratio of $d$ to the magnetic length $\ell$ is reduced experimentally,
the conductance peak appears to develop continuously starting at a critical 
value of $d/\ell$ that is consistent with earlier experimental anomalies\cite{Murphy}
attributed to spontaneous coherence and with mean-field-theory estimates of the 
critical layer separation \cite{Murphy} at which coherence is expected to develop.
These experiments are still not understood
quantitatively and raise a number of interesting issues in non-equilibrium
collective transport theory that have  
stimulated a growing body of theoretical \cite{bilayersrecent} work.

Since balanced bilayer systems at large odd integer total filling factor ($\nu_T \ge 9$)
are composed of 2D layers that, if isolated, would show stripe-state behavior, it 
is natural to consider the possible interplay and competition between
the formation of striped phases in each 2D layer and the development of 
spontaneous interlayer phase-coherence.  These issues have been investigated in several 
recent theoretical papers \cite{Brey,Cote,Demler} and it has been argued \cite{Demler} 
that they may be relevant for understanding a recent observation of 
resistance anisotropy at integer filling factor by Pan {\it et al.} \cite{Panintegeranisotropy}.
In the present paper we extend earlier work by two of the present 
authors \cite{MacDonald} on smectic states in single layers to the case
of bilayer systems.  The approach we take is one that is intended to be valid
when quantum fluctuation corrections to the stripe states predicted by
Hartree-Fock theory are weak on microscopic length scales, although as we discuss at length
they inevitably alter the ultimate physics at very low energies and temperatures and
the behavior of correlation functions at long distances.  Since stripe
states occur as extrema of the Hartree-Fock energy functional for any orbital
Landau level index, not only for $N \ge 2$ where the states are seen experimentally, 
and are in fact {\em always} unstable to the formation of Wigner crystal states in mean-field-theory, 
it is evident that we must appeal in part to experiment to judge when our starting
assumption is valid. \cite{Fogler,Koulakov,notbadthough}

In describing stripe states it is convenient to use a Landau gauge basis 
with single-particle states extended in the direction along the stripes, which
we choose to be the $\hat x$ direction, and 
labeled by a one-dimensional wavevector $k$ that is proportional to 
the guiding center along which the wavefunction's $y$-coordinate is localized, $Y = k \ell^2$.
For balanced bilayers, the stripe states that occur in Hartree-Fock theory are 
occupation number eigenstates in this representation, with the valence Landau-level
Landau gauge states occupied by top and bottom layer electrons in alternating stripes.
In the Hartree-Fock approximation, the low-energy excitations of the stripe states 
consist of coupled particle-hole excitations along each edge of top and bottom 
layer stripes. These degrees of freedom are 
conveniently described using the bosonization techniques familiar
from the theory of one-dimensional electron systems. \cite{Tsvelik}
Our approach is partly in the spirit of Fermi liquid theory 
in that we assume that the Hilbert space
of low-energy excitations can be placed in one-to-one correspondence with those that occur 
in the Hartree-Fock theory.  When quantum fluctuations are too strong our approach will not
be useful; for example, it cannot predict either the fact that lowest Landau level isolated layers
have composite-fermion liquid rather than stripe ground states, or the likelihood of 
{\em bubble}\cite{Koulakov} rather than stripe states far away from half-filling. 
Our approach to stripe state physics is similar to that taken first by 
Fradkin and Kivelson \cite{Fradkin}.  For the case of monolayers, the  
microscopic basis of the coupled Luttinger liquid model for quantum Hall stripe states was 
carefully examined by Lopatnikova {\em et al.} \cite{Lopatnikova} and other properties of 
quantum Hall stripe states have been addressed by Barci {\em et al.}\cite{Barci} and 
Wexler and Dorsey.\cite{Dorsey}    

Our paper is organized as follows.  In Section II we review the coupled 
Luttinger liquid model for quantum Hall stripe states and discuss its application to
the bilayer case.  The model rests fundamentally on the assumption that the 
excitation spectrum of bilayer stripe states may be placed in one-to-one 
correspondence with that of the Hartree-Fock picture; for bilayers this assumption
implies that the degrees of freedom at each stripe edge are those of a one-dimensional
electron gas.  Our analysis of the low-energy long-wavelength physics examines this 
subspace of the microscopic many-particle Hilbert space and includes {\em forward scattering}
terms in the Hamiltonian that create and destroy particle-hole excitations at the stripe 
edges, and {\em back scattering} terms that scatter electrons between chiral one-dimensional
electron-gas branches.  Since the microscopic amplitude of back-scattering
processes is weak, they can often be neglected at experimentally accessible temperatures.
When only forward scattering terms are included, the Hamiltonian can be
solved exactly using bosonization and is formally equivalent to that of a system of coupled
one-dimensional electron gases.  The quantum smectic broken symmetry 
character of the electronic state is reflected, however, in the coupled 
Luttinger liquid interaction parameters and results in  enhanced  
fluctuations.
The properties of this bilayer smectic state are discussed in Section III.
The behavior of the one-particle Green's functions at the smectic fixed point,
carefully addressed by Lopatnikova {\em et al.} \cite{Lopatnikova} for the 
single layer case, is discussed for the bilayer. 
We find that, as in the single-layer case, the one-particle Green's function does 
not exhibit the power law behavior that is generic for weakly coupled Luttinger liquids 
and instead vanishes faster than any power law at large distances, implying strongly
suppressed tunneling at low energies.   The enhanced importance of quantum fluctuations 
is a consequence of the 
invariance of the model's Luttinger liquid Hamiltonian under a simultaneous translation 
of all stripes.  Back-scattering interactions are addressed in Section IV, using a perturbative 
RG approach.  As in the single-layer
case we find that back-scattering interactions are always relevant. {\em The gapless 
Hartree-Fock smectic state
cannot be the true ground state in either single-layer or bilayer quantum Hall systems.}
Instead we conclude that except at relatively large layer separations, interlayer interactions 
induce a ground
state that has spontaneous interlayer phase coherence.  
This state would be signaled experimentally by the simultaneous occurrence of 
an integer quantum Hall effect and anisotropic finite-temperature transport, something
that has not been seen in single-layer systems.  Where intralayer
interactions are more important, they drive the system to a state with an anisotropic Wigner crystal in
each layer.  We argue that both types of interactions lead to charge gaps and to integer
quantum Hall effects and estimate the size of the resulting energy gaps.   According to 
our estimates, the gap created by inter-layer back-scattering will be large
enough to be observable out to surprisingly large layer separations.
The effect of finite tunneling between the layers is also addressed in Section IV.
Finally in Section V we discuss several interesting 
theoretical issues that arise from this work.  
We comment explicitly on inconsistencies between the conclusions that have been reached by 
different researchers on the question of smectic state stability in the single-layer
case.  We also address the suggestion\cite{Barci} that the enhanced quantum fluctuations that
follow from the broken translational symmetry of the starting Hartree-Fock state 
may invalidate our perturbative renormalization group analysis for back-scattering interactions.

\section{The model}

\subsection{Coupled Luttinger Liquid Model Energy Functional} 

\begin{figure}
\hspace{3mm}
\vspace{3mm}
\centerline{\includegraphics[width=8cm]{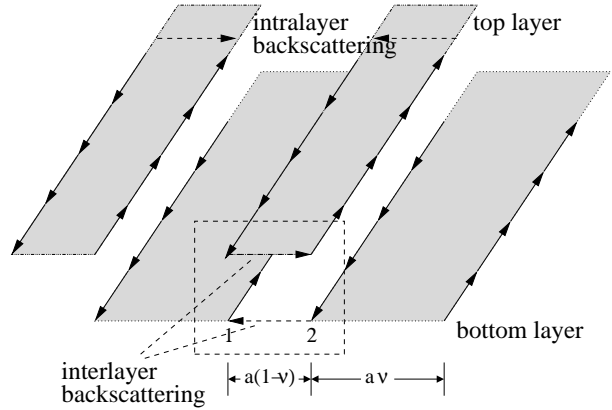}}
\caption{Schematic illustration of the Hartree-Fock bilayer smectic
state. The shaded areas are electron stripes whose edges are chiral
Luttinger liquids as denoted by the solid arrows. Each electron (hole) stripe
in one layer faces a hole (electron) stripe in the other.
The average filling factor of the highest occupied Landau level
is $\nu$ while that in the top layer is $1-\nu$, giving
a total filling factor $\nu_{T}=1$. In the convention used here, a rung
pair consists of the edges of an electron stripe in the top layer
and an hole stripe in the bottom layer. The right-moving and left-moving
chiral quasiparticle branches of each element of a rung pair are localized in opposite
layers and denoted by $\lambda=1,2$. The momentum-conserving
back-scattering interactions not present in the Luttinger liquid model,
discussed in this section, include both
interlayer and intralayer processes which have different behavior.
The figure illustrates interlayer and intralayer back-scattering processes
with the smallest possible momentum transfer.
\label{fig1}}
\end{figure}

In Hartree-Fock theory the smectic bilayer state at total filling
factor $\nu_{T}=1$ is a single Slater determinant where the occupation
of guiding-center modes in a Landau level of index $N\in\{0,1,2,\dots\}$
alternates between the layers with period $a$,
as depicted schematically in Fig.~\ref{fig1}.  Lower index Landau
levels are assumed to be frozen in filled states and higher index in empty states,
allowing them to be neglected in the following.  Each stripe has 
chiral left-moving
and right-moving branches of quasiparticle edge states, localized in opposite layers,
allowing the low-energy degrees of freedom of each electron stripe 
to be mapped to those of a one-dimensional electron gas. 
We consider the general case of a {\em biased} double layer system where the
width of the stripes in one layer is $a \nu$ while that in the other layer 
is $a(1-\nu)$, $\nu\in[0,1]$.  As indicated in the figure, for $\nu\neq 0.5$ 
the system has two types of stripe edges, distinguished by the direction 
of their closest neighbor.  In the following we refer to a pair
of chiral stripe edges one above the other in each layer
as a {\it rung}, and the two
closest such rungs as a {\it rung pair}.

Small fluctuations in the positions and shapes of the stripes can be described
in terms of particle-hole excitations near the stripe edges. The residual
interactions, ignored in Hartree-Fock theory, which act on 
energy states fall into two classes: {\em forward} scattering interactions which
conserve the number of electrons on each edge of every stripe, and {\em backward} 
scattering processes which do not.  
The quantum smectic model includes 
%
%
%
%
\hspace{-3.2mm}forward scattering processes only. 
Backscattering processes involve large momentum
transfer and their bare matrix elements will be smaller in
magnitude (see below).  We treat their effects perturbatively, using a 
renormalization group approach to account for the infrared divergences  
ubiquitous in quasi one-dimensional electron systems.  The smectic state is 
stable only if the back-scattering interactions are irrelevant. 

The {\em classical} quadratic Hamiltonian which describes the
energetics of small stripe edge fluctuations
has the following general form\cite{MacDonald}:
\begin{eqnarray}
{\cal H}_{0} & = & \frac{1}{2\ell^{2}}\int \rd x\int \rd x'
\sum_{j,k=-\infty}^{\infty}\sum_{\lambda,\mu=1,2}\sum_{\alpha,\beta=R,L}
\nonumber\\[3mm]
 & & \left[u_{j\alpha}^{\lambda}(x)K^{\lambda\mu}_{\alpha\beta}(x-x',j-k)
u_{k\beta}^{\mu}(x')\right] \quad ,
\label{eq:hamiltonian} 
\end{eqnarray}
where the indices $j$ and $k$ label rung pairs, $\lambda$ and $\mu$
the two different rungs within a rung pair, $\alpha$ and $\beta$
the right or left moving chiral edges within a single rung, and $x$, $x'$ positions
along the stripes.  In this equation, $u_{j \alpha}^{\lambda}$ is the transverse
displacement of the edge $(j,\lambda,\alpha)$ from its classical ground state location.

In Eq.~(\ref{eq:hamiltonian}), the linear charge density associated with an edge displacement is  
$\rho_{j \alpha}^{\lambda}(x)=\alpha n u_{j \alpha}^{\lambda}(x)$, where $n$ is the two-dimensional 
electron density inside the stripes $n= 1/2\pi l^2$ and $\alpha=-,+,$ or $L,R,$ for the left, right, 
moving fermions respectively.  It follows from symmetry considerations that the elastic kernel 
satisfies the following equalities,
\bea
K_{\alpha \beta}^{\lambda \mu}(j)=K_{\beta \alpha}^{\lambda \mu}(j)
=K_{(-\alpha) (-\beta)}^{\lambda \mu}(j)=K_{\alpha \beta}^{\lambda \mu}(-j) \quad ,
\eea
which allow the Hamiltonian to be rewritten into sums and differences of the 
positions of left and right-going branch edges:
{\small
\begin{eqnarray}
\label{symm-asymm}
{\cal H}_{0} & = & \frac{1}{8 l^4}\int \rd x\int \rd x'
\sum_{j,k=-\infty}^{\infty}\sum_{\lambda,\mu=1,2} \biggl[ \biggr.
\\[3mm]\nonumber
& &\hspace{-8mm} \left[u_{j,R}^{\lambda}(x)+{u}_{j,L}^{\lambda}(x)\right] 
K^{\lambda\mu}_{\rm p}(x-x',j-k)
\left[u_{k,R}^{\mu}(x')+{u}_{k,L}^{\mu}(x')\right]+ 
\\[3mm]\nonumber
& &\hspace{-8mm} \left[u_{j,R}^{\lambda}(x)-{u}_{j,L}^{\lambda}(x)\right]
K^{\lambda\mu}_{\rm c}(x-x',j-k)
\left[u_{k,R}^{\mu}(x')-{u}_{k,L}^{\mu}(x')\right]  \biggl. \biggr]
\; ,
\end{eqnarray} }
where 
\bea
K^{\lambda\mu}_{\rm p}(x,j)&=&\ell^{2}\sum_{\alpha\beta}K_{\alpha\beta}^{\lambda\mu}(x,j)
\quad ,
\\[3mm]
K^{\lambda\mu}_{\rm c}(x,j)&=&\ell^{2}\sum_{\alpha\beta}
\alpha\beta K_{\alpha\beta}^{\lambda\mu}(x,j) \quad .
\eea
We regard the first term in this Hamiltonian as the contribution from fluctuations
in the {\em position} of the rungs while the second term is the contribution from
fluctuations of their charge densities.  (The total filling factor varies locally when 
left and right going branch edges do not move together.)  These two terms are 
analogous respectively to current and charge terms in the effective Hamiltonian 
of a conventional one-dimensional electron gas.
The calculations we perform here will require only the long-wavelength limits of
the $x-x'$ dependence of elastic kernel in this Hamiltonian, which we estimate
using a weak-coupling approximation that we discuss below.  


\subsection{Bosonization} 

This Hamiltonian is quantized by recognizing that charge and position
fluctuations result from particle-hole excitations at the edges of 
chiral quasiparticle branches, just as in an ordinary one-dimensional electron
system. The real spin is frozen due to the presence of strong perpendicular 
magnetic field and as a result we bosonize according to spinless bosonization 
scheme\cite{Tsvelik}.  It follows from standard arguments that 
\begin{eqnarray}
\left[\rho_{j,\alpha}^{\lambda}(x),\rho_{k,\beta}^{\mu}(x')\right]
&=&\frac{i}{2\pi}\delta_{\lambda,\mu}
\delta_{\alpha,\beta}\delta_{j,k}\partial_{x}\delta(x-x')\quad .
\end{eqnarray}
In terms of Fermion creation and annihilation operators 
{\small
\bea
\rho_{jR}^{\lambda}(x) & = & :R_{j}^{\lambda \dagger}(x)R_{j}^{\lambda}(x):
=R_{j}^{\lambda \dagger}(x)R_{j}^{\lambda}(x) - 
\left<R_{j}^{\lambda \dagger}(x)R_{j}^{\lambda}(x) \right> \, , 
\\ [3mm] 
\rho_{jL}^{\lambda}(x) & = & :L_{j}^{\lambda \dagger}({x})
L_{j}^{\lambda}({x}): = L_{j}^{\lambda \dagger}({x}) L_{j}^{\lambda}({x})
-\left< L_{j}^{\lambda \dagger}({x}) L_{j}^{\lambda}({x}) \right>\, ,
\eea
}
with $\lambda\in\{1,2\}$ denoting the rung in rung pair $j$ and $R$, $L$,
labeling right and left movers at the stripe edges.

The low energy Hamiltonian is more conveniently described in terms of 
boson fields.  The  right and left moving fermionic fields on the 
left stripe edge of rung pair $j$ are given by 
\bea
{\psi^1_j}_{R}(x)=e^{i[b(j-1/2) - k_{\rm F}]x}R^1_j(x)
\label{fermion_1} \quad ,
\eea
while those on the right are given by, 
\bea
{\psi^2_j}_{R}(x)=e^{i[b(j-1/2) + k_{\rm F}]x}R^2_j(x) \quad .
\label{fermion_2}
\eea
The above equations hold similarly for the left movers with the only change
$R\rightarrow L$.
Here $b=a/l^2$ is the width in $k$ space of a rung and $k_{\rm F}= a\nu /2l^2$ is the Fermi wavevector
for the bottom layer stripes.
The right and left slow fields can be expressed in terms of boson fields as in 
conventional one-dimensional electron systems: 
\begin{equation}
R_j^\lambda(x)=\frac{1}{\sqrt{2\pi}}e^{i\phi_{j,R}^\lambda(x)} \quad , \quad 
L_j^\lambda(x)=\frac{1}{\sqrt{2\pi}}e^{i{\phi}_{j,L}^\lambda(x)} \quad , \quad
\end{equation}
where $\phi_{j,R}^\lambda(x)$ and ${\phi}_{j,L}^\lambda(x)$ are the chiral
 components of the bosonic field
$\Phi_j^\lambda(x)=\left[\phi_{j,R}^\lambda(x)+{\phi}_{j,L}^\lambda(x)\right]/2$.

In terms of the bosonic fields the chiral currents take the following form 
\bea
\rho_{j \alpha}^{\lambda}(x)  =  -\frac{\alpha}{2\pi}\partial_{x} 
\phi^{\lambda}_{j,\alpha}(x) \quad .
\eea
Introducing the dual field $\Theta_j^\lambda(x) = \hspace{-0.9mm}
\left[\phi_{j,R}^\lambda(x) - {\phi}_{j,L}^\lambda(x)\right]/2$,
the position and charge variables $U_j^\lambda$, ${U'}_j^\lambda$, of the two edge system
can be expressed as $U_j^\lambda = - l^2 \p_x \Phi_j^\lambda$ and
${U'}_j^\lambda = - l^2 \p_x \Theta_j^\lambda$. This shows that the field $\Phi$ is related to 
the position fluctuations of the two edge system whereas $\Theta$ is related with their
charge density fluctuations.


The field $\Phi$ and the dual field $\Theta$ satisfy 
the following commutation relation 
\bea
\left[\Theta_j^\lambda(x), \p_{x'} \Phi_k^\mu\right(x')] = 
-i \pi \delta_{j,k} \delta_{\lambda,\mu} \delta(x-x') \quad .
\eea
(The fields $\Theta/\sqrt{\pi}$ and $-\p_x\Phi/\sqrt{\pi}$ are canonical conjugates).
In terms of these new fields the Hamiltonian takes the following form
\bea
{\cal H}_0&=&
\frac{1}{2} \int \hspace{-1.2mm}
\rd x \rd x' \sum_{j,k} \hspace{-1.0mm} \left[ \p_{x} \Phi_j^\lambda(x)
K_\Phi^{\lambda \mu}(x-x',j-k)\p_{x'}\Phi_k^\mu(x') +
\right.
\nonumber\\[3mm]
&& \left. \hspace{5mm}
+ \p_{x} \Theta_j^\lambda(x)
K_\Theta^{\lambda \mu}(x-x',j-k) \p_{x'} \Theta_k^\mu (x')
\right]
\quad ,
\eea
whereas the Fourier transform of the corresponding action 
 \begin{equation}
 {\cal S}_{0}=\frac{i}{\pi}\int {\rm d}x {\rm d}\tau\sum_{j}
 \sum_{\lambda}
 \left(\partial_{x}\Phi_{j}^{\lambda}\right)\partial_{\tau}
 \Theta_{j}^{\lambda}+\int d\tau{\cal H}_{0}
 \end{equation}
 reads
\begin{eqnarray}
{\cal S}_{0} & = & \int_{{\bf q},\omega}
\Biggl[\sum_{\lambda}\left(\frac{i}{\pi}q_{x}
{\Phi^{\lambda}}^{*}({\bf q},\omega)\omega\Theta^{\lambda}({\bf q},\omega)\right)
\nonumber\\[3mm]
& & +\frac{1}{2}\sum_{\lambda,\mu}\biggl(
q_{x}^{2}{\Phi^{\lambda}}^*({\bf q},\omega)K_{\Phi}^{\lambda\mu}({\bf q})
\Phi^{\mu}({\bf q},\omega)\nonumber\\[3mm]
& &\qquad\qquad
+q_{x}^{2}{\Theta^{\lambda}}^*({\bf q},\omega)K_{\Theta}^{\lambda\mu}({\bf q})
\Theta^{\mu}({\bf q},\omega)\biggr)\Biggr] \quad . 
\label{actionphitheta}
\end{eqnarray}
Here we have employed the shorthand notation,
\begin{equation}
\int_{{\bf q},\omega}=\int_{-\Lambda}^{\Lambda}\frac{\rd q_{x}}{2\pi}
\int_{-\frac{\pi}{a}}^{\frac{\pi}{a}}\frac{\rd q_{y}a}{2\pi}
\int_{-\infty}^{\infty}\frac{d\omega}{2\pi}
\label{integr_bound}
\end{equation}
with $\Lambda \sim 1/\ell$ a high momentum cutoff, and have adopted
the 
following Fourier transform conventions:
\begin{eqnarray}
F_{j}(x,\tau) & = & \int_{{\bf q},\omega}e^{i(q_{x}x+q_{y}aj-\omega\tau)}
F({\bf q},\omega)\quad ,
\\[3mm]
F({\bf q},\omega) & = & \int \rd x \rd \tau\sum_{j}
e^{-i(q_{x}x+q_{y}aj-\omega\tau)}
F_{j}(x,\tau)\quad .
\end{eqnarray}
The QFT applies for distances larger than $l$ and as a result the 
$q_x$ integration has to be cut off by $\pm 2\pi/l$.
In Eq.~(\ref{actionphitheta}) the kernel matrices 
$K_{\Phi}({\bf q})$ and $K_{\Theta}({\bf q})$
are the Fourier transforms of
\begin{eqnarray}
K_{\Phi}^{\lambda\mu}(x,j)=K_{\rm p}^{\lambda\mu}(x,j)
& = & 2\ell^{2}\hspace{-1mm}\left[K_{RR}^{\lambda\mu}(x,j)+K_{RL}^{\lambda\mu}(x,j)\right]
\, ,
\label{defKphi}\\[4mm]
K_{\Theta}^{\lambda\mu}(x,j)=K_{\rm c}^{\lambda\mu}(x,j)
& = & 2\ell^{2}\hspace{-1mm}\left[K_{RR}^{\lambda\mu}(x,j)
-K_{RL}^{\lambda\mu}(x,j)\right]\, .
\label{defKtheta}
\end{eqnarray}
Integration over the $\Phi$ fields in (\ref{actionphitheta}) yields an 
effective action in terms of the $\Theta$ fields alone
\begin{eqnarray}
{\cal S}_{\Theta} & = & \frac{1}{2}\int_{{\bf q},\omega}
\sum_{\lambda,\mu}\biggl[\Theta^{\lambda *}({\bf q},\omega)
\nonumber\\[3mm]
 & & \cdot\left(\frac{\omega^{2}}{\pi^{2}}
\left(K^{-1}_{\Phi}({\bf q})\right)^{\lambda\mu}
+q_{x}^{2}K_{\Theta}^{\lambda\mu}({\bf q})\right)
\Theta^{\mu}({\bf q},\omega)\biggr]\quad .
\label{thetaaction}
\end{eqnarray}
The corresponding ${\cal S}_{\Phi}$ action, obtained by
integrating out the $\Theta$ fields, differs only through the interchange of 
$\Phi$ and $\Theta$, and $K_{\Phi}$, $K_{\Theta}$.

\subsection{Microscopic Theory of Long-Wavelength Interaction Parameters} 

The objective of this coupled Luttinger liquid model for stripe states in 
quantum Hall bilayers is to address the consequences of weak quantum fluctuations when the ground state 
is similar to the mean-field-theory stripe state.  In this spirit, we use weak-coupling
expressions for the interaction parameters of the model, replacing scattering amplitudes
by the bare values for scattering of the Hartree-Fock theory quasiparticles.
If the true ground state were a smectic, the values of these parameters would be 
renormalized somewhat by higher order scattering processes.  We expect that it will
prove difficult to systematically improve on the estimates given here for the quantum Hall bilayer 
case because of the absence of a one-body kinetic energy term in the relevant microscopic
Hamiltonian that would enable a systematic perturbative expansion.  
We emphasize that a quantitative theory of the forward scattering amplitudes that 
has a sound microscopic foundation 
is {\em necessary} in order to decide on the relevance of the back-scattering interactions 
we have neglected so far and the character of the true ground state.
As emphasized by the work of Fradkin, Kivelson and co-workers\cite{Fradkinold}
any conclusion is possible if the forward scattering interactions are allowed to vary
arbitrarily.  The perturbative renormalization group scaling dimensions
that we evaluate below are dependent only 
on the elastic constants at $q_{x}=0$, {\it i.e.} for strait stripe edges.
The weak-fluctuation Hamiltonian may be evaluated in this limit by calculating the 
expectation value of the microscopic Hamiltonian in the Hartree-Fock theory 
ground state, which in this limit is a single Slater determinant
with straight stripe edges displaced from those in the Hartree-Fock theory stripe 
ground state.  By evaluating the expectation value of the Microscopic Hamiltonian in a
state with arbitrary stripe edge locations we find that for 
$j\neq 0$:
\bea
\nonumber
&& \int dx K_{\Phi/\Theta}(x,j) = 
\\[3mm]
&& =\frac{1}{2\pi^2\ell^2}
\left( \begin{array}{cc}
V(ja)\mp W(ja) \;& W \mp V(ja-a\nu)\\ \\
W \mp V(ja+a\nu)\; & V(ja)\mp W(ja)
\end{array}\right) \quad .
\label{HFrealspaceK12}
\eea
In the off diagonal elements the argument of $W$ is the same as that of $V$,
$(ja \pm a\nu)$, respectively.
The $V$ and $W$ contributions are proportional to two-particle intralayer and interlayer 
interaction matrix elements respectively, and are given by 
\begin{eqnarray}
V(Y) & = & \int\frac{\rd q}{2\pi}e^{-\frac{1}{2}q^{2}\ell^{2}}
V_{S}^{N}(q)e^{-iqY}-e^{-\frac{1}{2}Y^{2}/\ell^{2}}
\nonumber\\ [3mm]
&  & \cdot\int\frac{\rd q}{2\pi}e^{-\frac{1}{2}q^{2}\ell^{2}}
V_{S}^{N}\left(\sqrt{Y^{2}/\ell^{4}+q^{2}}\right)\quad ,
\\[4mm]
W(Y) & = & \int\frac{\rd q}{2\pi}e^{-\frac{1}{2}q^{2}\ell^{2}}
V_{D}^{N}(q)e^{-iqY}\quad .
\label{W}
\end{eqnarray}
Note that the intralayer interaction contributions have competing direct and 
exchange contributions that cancel for $Y=0$ whereas the interlayer interaction has only a direct 
contribution. In the following we shall assume infinitely narrow quantum wells in both layers so that 
the interaction potentials occurring in the above equations read
\begin{equation}
V^{N}_{S/D}(q)=
\left[L_{N}\left(\frac{1}{2}q^{2}\ell^{2}\right)\right]^{2}V^{0}_{S/D}(q)\quad ,
\end{equation}
where $L_{N}(x)$ 
is the Laguerre polynomial form factor for electrons in the 
$N$-th excited Landau level, and $V^{0}_{S/D}$ is the Fourier transform of the Coulomb interaction
within and between the layers, $V_{S}^{0}(q)=e^{2}2\pi/|q|$,
$V_{D}^{0}(q)=e^{2}(2\pi/|q|)\exp(-|qd|)$ with $d$ being the layer separation.
The long-ranged nature of the Coulomb interaction  
leads to logarithmic divergences in $V$ and $W$ which we regularize by adding a
term $-(e^{2}2\pi/|q|)\exp(-2|q|d_{\rm gate})$ to $V^{0}_{S/D}$ with
$d_{\rm gate}\gg d$. This regularization can be roughly thought of as 
introducing a metallic screening plane at distance $d_{\rm gate}$ leading to image
charges that screen interactions between electrons in the bilayer system.
Although $V$ and $W$ diverge for $d_{\rm gate} \to \infty$, it is possible to
show\cite{singlelayer} that $K_{\Phi}$ and $K_{\Theta}$ remain finite.  
In the following 
we choose a large but finite value for $d_{\rm gate}$ for numerical convenience.   

The above form of the smectic energy kernel 
$K_{\Phi/\Theta}(j)$ applies for $j\neq 0$. For $j=0$ the components
$K_{RL}(0)=K_{LR}(0)$ and of $K_{RR}^{\lambda\mu}(0)=K_{LL}^{\lambda\mu}(0)$
for $\lambda\neq\mu$ are given by the same expressions. 
The quantities $K_{RR}^{11}(0)=K_{LL}^{11}(0)$ [$=K_{RR}^{22}(0)=K_{LL}^{22}(0)$] 
have additional contributions that originate from the wavevector dependence of the Hartree-Fock
self-energy at a given stripe edge and capture the key property that the energy of the smectic
must be invariant under rigid translations of all stripes,
$u_{j\alpha}^{\mu}(x)\mapsto u_{j\alpha}^{\mu}(x)+{\rm constant}$
\cite{MacDonald,Chaikin}. We find that 
\begin{eqnarray}
K_{RR}^{11}(0) & = & -\biggl[K_{RR}^{12}(q_{y})+K_{RL}^{12}(q_{y})
+K_{RL}^{11}(q_{y})\biggr]_{q_{y}=0}
\nonumber\\[3mm]
 & & -\sum_{j\neq0}K_{RR}^{11}(j)\quad .
\end{eqnarray}
Note that these properties imply that ${\rm det}[K_{\Phi}(q_y=0)]=0$.
When these long wavelength approximations are employed,
the Fourier transforms of $K^{\lambda \mu}_{\Phi/\Theta}$ 
in (\ref{HFrealspaceK12}) 
depend only on $q_{y}$ but not on $q_{x}$. From the relation
$K_{\Phi/\Theta}(x,-j)=K_{\Phi/\Theta}^{T}(x,j)$ it follows that
$K_{\Phi/\Theta}(q_{y})$ is hermitian. Under particle-hole
transformation, $\nu\mapsto 1-\nu$, the diagonal elements of these matrices
remain unchanged while the off-diagonal elements transform as
\begin{eqnarray}
K_{\Phi/\Theta}^{12}(q_{y}) & \mapsto & e^{-iq_{y}a}
K_{\Phi/\Theta}^{21}(q_{y}) \quad ,
\\[3mm]
K_{\Phi/\Theta}^{21}(q_{y}) & \mapsto & e^{+iq_{y}a}
K_{\Phi/\Theta}^{12}(q_{y}) \quad .
\end{eqnarray}

\subsection{The Balanced Bilayer Limit} 

The special case of half filling in each layer has additional symmetry that is most 
conveniently exploited by taking a slightly different approach.
In this case the electron stripes in both layers have the same width and the
system therefore has effective periodicity of $a/2$.  To be more precise, the 
problem can be formulated as a one-dimensional lattice of equidistant
double edges placed a distance $a/2$ apart, noting carefully that right and left
goers interchange their layer labels on alternate edges. 
To describe this system instead of using coupling matrices 
$K^{\lambda \mu}_\Phi(x,j)$ and $K^{\lambda \mu}_\Theta(x,j)$, as in the unbiased case,
one can simply use the coupling constants $\tilde{K}_\Phi(x,j)$ and $\tilde{K}_\Theta(x,j)$
\bea
\tilde{K}_{\Phi/\Theta}(x,j)    &=&  2 l^2 \left[K_{RR}(x,j) \pm  K_{RL}(x,j)\right] \quad ,
\eea 
with the value for $K_{RR}(0)$ reflecting translation invariance
\bea
K_{RR}(0)=-\frac{1}{4\pi^2 l^2}\sum_{j=-\infty}^{+\infty} (-1)^j\left[V(j)-W(j)\right] \quad .
\eea
In momentum space $\tilde{K}_\Phi$ and $\tilde{K}_\Theta$ have the following form
\bea
\tilde{K}_\Phi(q_y) & = & \frac{1}{2\pi^2}\left[V(q_y-\frac{2\pi}{a})-V(\frac{2\pi}{a})
\right.
\nonumber\\[0mm]
&& \hspace{17mm} \left. -W(q_y-\frac{2\pi}{a})+W(\frac{2\pi}{a})\right] \quad ,
\label{Kf-qy}
\\[3mm]
\tilde{K}_\Theta(q_y) & = &\frac{1}{2\pi^2} \left[V(q_y)-V(\frac{2\pi}{a})
+W(q_y)+W(\frac{2\pi}{a})\right] \,.
\label{Kth-qy}
\eea
We have used this simpler and partially independent formulation of the $\nu=1/2$ limit, 
to test our results for the general case.

\section{Smectic State Properties} 

A peculiar property of quantum Hall stripe states is that the microscopic scale of back-scattering
interactions is weak.  For this reason observable properties may be those of smectic states over
a wide interval of temperature, even when back-scattering interactions are relevant at the smectic 
fixed point.  In this section we discuss some characteristic properties of quantum Hall bilayer 
stripe states. 

\subsection{Collective modes and Thermodynamic Properties} 

The coupled Luttinger liquid model for bilayer quantum Hall stripe states 
gives rise to two collective modes with dispersions that can be determined by evaluating
zeros of the determinant of the $2 \times 2$ matrix that defines
%
%
\begin{figure}
\unitlength=1mm
\begin{picture}(80,29)
\put(-1.5,3){\epsfig{file=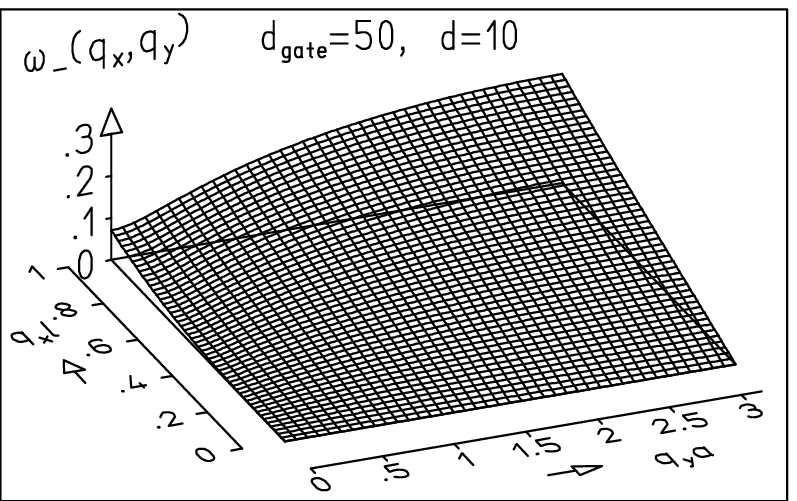,height=25mm}}
\put(38.5,3){\epsfig{file=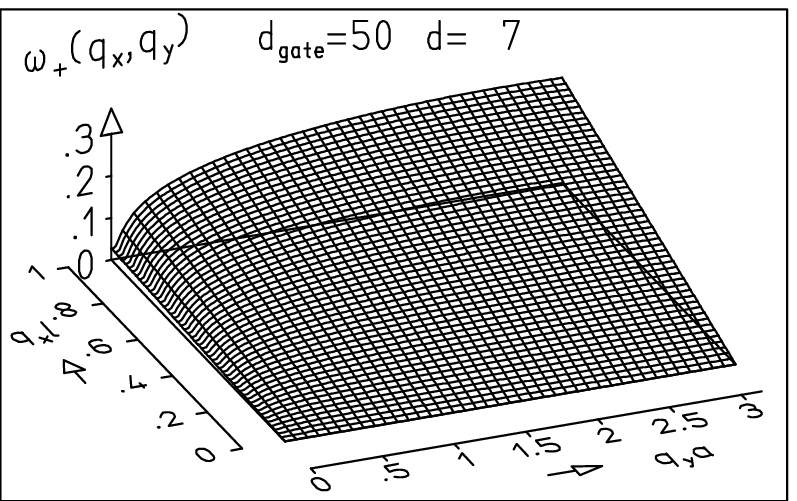,height=25mm}}
\end{picture}
\caption{Collective modes of the bilayer QH smectic phase.
These results shown here were evaluated using
$K_{\Phi/\Theta}(x,j)$ values from Eqs.~(\ref{HFrealspaceK12}-\ref{W}) with
$N=2$, $\nu\approx1/2$ and $a=5.8 l$.
Notice that $\omega_{+}(q_x,q_y)$ always disperses linearly for small
$q_x$, whereas $\omega_{-}(q_x,q_y)$
disperses sub-linearly (as $q_x^3$) when $q_y=0$.  The $q_y=0$ behavior
of the lower collective mode is sensitive to the $q_x$ dependence of
the interaction coefficients which we do not evaluate microscopically.
This illustration was constructed by adding a small $q_x^4$ contribution to
the interaction coefficients.
}
\label{modes}
\end{figure}
{\hspace{-3.2mm}the real time quadratic action
at each $\bf q$ and $\omega$.  Writing this matrix (with indices suppressed) 
as $ K_{\Phi}^{-1} * [\omega^2 +  \pi^2 q_x^2 K_{\Phi} K_{\Theta}]/\pi^2$, it follows that 
the squares of the quadratic boson collective mode energies are 
\bea
\omega^2_{\pm}({\bf q}) &=& v^2_{\pm}({\bf q}) q^2_x =
\pi^2 q^2_x \frac{ {\rm Tr}\left[K_\Phi({\bf q}) K_\Theta({\bf q})\right]}{2}\cdot
\nonumber\\[3mm]
&&\cdot\left[1\pm
\sqrt{1-\frac{4\det\left[K_\Phi({\bf q}) K_\Theta({\bf q})\right]}
{ {\rm Tr}^2\left[K_\Phi({\bf q}) K_\Theta({\bf q})\right]} }\right] \quad .
\label{modes0}
\eea
Both modes have energies that are proportional to $q_x$. 
The velocity of the $\omega_-({\bf q})$ mode vanishes for $q_y \to 0$.
In fact, when the $q_x$ dependence of  $K_{\Phi}$ and $K_\Theta$ is dropped, as 
in most of our calculations $\omega_-(q_x,q_y=0)$ vanishes identically; 
when the $q_x$ dependences are restored $\omega_-^2({\bf q}) \approx q_x^2(q_y^2+ q_x^4)$ 
at small wavectors and $\omega_-(q_x,q_y=0) \propto q_x^3$. 
For gate-screened Coulomb interactions, the x-direction $\omega_-$ mode velocity is
proportional to $|q_y|$ in the small $q_x$ and $q_y$ limit. 
In the independent layer limit, the two modes become degenerate and 
we recover the isolated layer results obtained previously \cite{MacDonald}.

In the case of balanced bilayers the alternate formulation mentioned above is
more convenient. The collective modes for this limit may be expressed as 
\bea
\omega_{1,2}({\bf q})=
\pi q_x \sqrt{\tilde{K}_\Phi({\bf q}) \tilde{K}_\Theta({\bf q})} \quad ,
\label{colmodehalf}
\eea
where $\tilde{K}_\Phi({\bf q})$ and $\tilde{K}_\Theta({\bf q})$ are given by 
Eqs.~(\ref{Kf-qy},\ref{Kth-qy}).  The two collective modes 
of the general formulation applied to the $\nu = 1/2$ case
correspond to two different wavevectors of this dispersion relation. 
The collective modes of the bilayer QH smectic phase are shown in Fig.~\ref{modes}. 
The right panel shows $\omega_{+}(q_x,q_y)$ which disperses
linearly in small $q_x$ for arbitrary $q_y$.  In contrast,
$\omega_{-}(q_x,q_y)$ disperses linearly at small $q_x$
for $q_y \ne 0$, but for $q_y=0$ it is sub-linear:
$\omega_{-}(q_x,q_y=0) \sim q_x^3$.  

The Thermodynamic properties of the smectic phase of the bilayer system 
are those of a non-interacting boson system and are readily evaluated given
the collective mode energies.  
For example, for $\nu=1/2$ in each layer, using the simpler alternative formulation, we 
have one collective mode at each wavevector.  The internal energy density is
\bea
U=\int \frac{d^2 q}{(2\pi)^2/a} \frac{\omega(\bf{q})}{e^{\omega(\bf{q})/T}-1} \quad ,
\eea
where $\omega(\bf{q})$ is given by Eq.~(\ref{colmodehalf}). 
At low temperatures only the long-wavelength behavior matters and we obtain 
\bea
U & \approx & \int \frac{d q_x \rd q_y}{(2\pi)^2/a} 
\frac{ \pi |q_x| |q_y| \sqrt{\tilde{K_\theta}(0) \tilde{K}_\Phi^{''}(0)}} {
e^{\pi |q_x| |q_y| \sqrt{\tilde{K_\theta}(0) \tilde{K}_\Phi^{''}(0)}/T}-1} 
\nonumber\\[3mm]
& = &
\frac{2a}{\pi^3 \sqrt{\tilde{K_\theta}(0) \tilde{K}_\Phi^{''}(0)} } 
T^2 \zeta(2)\ln\left(\frac{T_0}{T}\right) \quad ,
\label{U_unbiased}
\eea
where $k_B T_0 = \sqrt{\tilde{K_\theta}(0) \tilde{K}_\Phi^{''}(0)}/al$.
The specific heat of this system varies as $T \ln(T_0/T)$ at small $T$, vanishing 
less quickly than that of a non-interacting Fermion system because of the 
soft collective modes that result from the translational invariance of the stripe state. 
This low temperature behavior reflects the form of the dispersion relation at small $\bf{q}$, 
$\omega_{-}({\bf q})\sim q_x q_y$; only the prefactor of this enhanced specific heat 
changes in the unbalanced bilayer case.  These results for the specific heat are
similar to the ones obtained in previous works by Barci {\it et al.}\cite{Barci}, and 
Lopatnikova {\it et al.} \cite{Lopatnikova} for the case of a single layer.  
There is no qualitative difference between the 
thermodynamic properties of single layer and bilayer stripe states.
For unbalanced bilayers the specific heat at low temperatures is dominated by the 
softer of the two collective modes, whose long wavelength dispersion is given
by 
\bea
\omega_-({\bf q})\approx q_x q_y \sqrt{\frac{{\rm Tr}\left[K_\Phi(0)K_\theta(0)\right]}{
\det[K_\theta(0)] [\det(K_\Phi)]^{''}(0)}} \quad .
\eea
It follows that the internal energy is given by  
\bea
U  \approx 
	\frac{a}{\pi^3} \sqrt{\frac{{\rm Tr}\left[K_\Phi(0)K_\theta(0)\right]}{\det(K_\theta(0)) [\det(K_\Phi)]^{''}(0)} }
	T^2 \zeta(2)\ln\left(\frac{T_0}{T}\right) \quad ,
\label{U_biased}
\eea
and the specific heat will  vary again as $T \ln(T_0/T)$.
The $T_0$ in Eq.~(\ref{U_biased}) and Eq.~(\ref{U_unbiased}) are given by corresponding expressions.

\subsection{Boson and Fermion correlation functions at the smectic fixed point}

In this section we discuss the static and dynamic correlation functions
of the right and left moving fermions of the stripe edges and the Boson correlation
functions in terms of which they are evaluated. 
The right and left moving fermion fields are expressed in terms of the 
$\Phi$ and $\Theta$ boson fields as follows:
\bea
R_j^\lambda(x,\tau)&=&\frac{1}{\sqrt{2\pi}} e^{i\phi_{j,R}^\lambda(x,\tau)}=
\frac{1}{\sqrt{2\pi}} 
e^{i\left[\Phi_j^\lambda(x,\tau)
+\Theta_j^\lambda(x,\tau)\right]}\; ,
\\[3mm]
L_j^\lambda(x,\tau)&=&\frac{1}{\sqrt{2\pi}} e^{i{\phi}_{j,L}^\lambda(x,\tau)}
=\frac{1}{\sqrt{2\pi}}
e^{i\left[\Phi_j^\lambda(x,\tau) - \Theta_j^\lambda(x,\tau)\right]}
\; ,
\eea 
where the $\Phi$ field is related to position fluctuations of the two edge system, 
while the $\Theta$ field is related to its charge density fluctuations.
We observe in the following that the charge and position fluctuations of the edges have a 
dramatically different effect on the correlation functions of the right and left movers.
The single-particle Green's function for the right movers is given by
\bea
&&\left<{R_j^{\lambda}}^\dagger(x,\tau)R_j^{\lambda}(0,0)\right>= 
\\[3mm]\nonumber
&&\hspace{0mm}= \frac{1}{2\pi}
e^{-\frac{1}{2}\left<\left[\Phi_j^\lambda(x,\tau)-\Phi_j^\lambda(0,0)\right]^2\right>}
e^{-\frac{1}{2}\left<\left[\Theta_j^\lambda(x,\tau)-\Theta_j^\lambda(0,0)\right]^2\right>}
\quad .
\eea
We first evaluate the $\Phi$ and $\Theta$ field correlation functions
$\tilde{C}^{\lambda \lambda}_{\Phi}$ and $\tilde{C}^{\lambda \lambda}_\Theta$, where
\bea
\tilde{C}^{\lambda \lambda}_\Phi(x,0,0) &=& \left<\left[\Phi_j^\lambda(x,\tau)
-\Phi_j^\lambda(0,\tau)\right]^2\right> \quad ,
\label{tilde_C}
\eea
and similarly for $\tilde{C}^{\lambda \lambda}_\Theta$.
In Eq.~(\ref{tilde_C}) and in the following the arguments of 
$\tilde{C}^{\lambda \lambda}_{\Phi/\Theta}$ are $(x-x',j-j',\tau-\tau')$.
From Eq.~(\ref{thetaaction}) we have for the $\Phi_j^\lambda$ field
\bea
\tilde{C}^{\lambda \lambda}_\Phi(x,0,0) 
=2\int\frac{\rd^2 q \rd \omega}{(2\pi)^3/a} \left[1 -\cos(q_x x)\right]
\left[M_{\Phi}^{-1}({\bf q},\omega)\right]^{\lambda \lambda}  ,
\eea
where
\bea
M_{\Phi}^{\lambda \mu}({\bf q},\omega)=\frac{\omega^2}{\pi^2}\left[K_\Theta^{-1}(q_y)\right]^
{\lambda \mu} + q_x^2 \left[K_\Phi(q_y)\right]^
{\lambda \mu}
\quad .
\label{M_matrix}
\eea
$M_\Theta({\bf q},\omega)$ can be obtained by interchanging $K_\Phi$, $K_\Theta$.
The integral over $\omega$ is readily valued by decomposing 
$M_{\Phi}^{-1}$ as a sum over eigenmode contributions, writing it in the
form $\sum_{\pm} C^{\lambda \mu}_{\pm}/\left[\omega^2 + \omega_{\pm}^2(\bf q)\right]$.
It follows after some algebra that correlation function can be expressed 
in the form  
\bea
\tilde{C}^{\lambda \lambda}_\Phi(x,0,0)&=&\int_0^\Lambda \rd q_x \frac{\left[1-\cos(q_x x)\right]}{q_x}
\int_{-\pi/a}^{\pi/a} \frac{\rd q_y}{2\pi/a}\cdot 
\nonumber\\[3mm]
&&\cdot\left[\frac{\det(K_\Theta)}{K_{+} + K_{-}}
\left[(K_\Theta^{-1})^{\lambda \lambda}+
\frac{K_\Phi^{\lambda \lambda}}{K_{+} K_{-}}\right]\right] 
\label{Phi-correl}
\\[3mm]\nonumber
&& \hspace{-20mm}\approx  
\ln(\frac{|x|}{l}) \int \frac{\rd q_y}{2\pi/a} 
\left[\frac{\det(K_\Theta)}{K_{+} + K_{-}}
\left[(K_\Theta^{-1})^{\lambda \lambda}+
\frac{K_\Phi^{\lambda \lambda}}{K_{+} K_{-}}\right]\right] \, ,
\eea
for large $x$, where $K_{\pm} = v_{\pm}(\bf q)/\pi$. 
In the limit of balanced filling fraction for which $K_{\Phi}$ and $K_{\Theta}$ are scalars 
the integrand which is averaged over
$q_y$ above reduces to $\sqrt{{\tilde K}_{\Theta}/{\tilde K}_{\Phi}}$, the familiar result for $\Phi$-field
correlation functions in a standard one-dimensional electron system.  This result is generalized
here by the average over $q_y$ and by the particular way in which the matrix nature of the 
$K_{\Theta}$ and $K_{\Phi}$ expressions enter the matrix elements above.   The result for the 
$\Theta$ field correlation functions differs only by the interchange of the $K_{\Theta}$ and 
$K_{\Phi}$ matrices.  At first sight it appears that the position fluctuation factor in 
the right mover correlation 
function decays algebraically along the stripes. However, the power which characterizes 
this decay, $d_{\Phi,x}$, is 
given by the integral over $q_y$ of Eq~(\ref{Phi-correl}), which diverges logarithmically
because $K_{-} \propto v_{-}$ vanishes as $|q_y|$ as $|q_y| \to 0$; the same soft position 
fluctuations that led above to an enhanced specific heat, lead here to fermion correlation functions
that decay faster than any power low but slower than an 
exponential. This observation generalizes to bilayers, a property of single layer stripe
states noted by Lopatnikova {\em et al}\cite{Lopatnikova} and 
Barci {\em et al}\cite{Barci}. 

$\tilde{C}_\Theta(x,0,0)$, which specifies the charge fluctuation factor in the 
fermion correlation functions, is given by
\bea
&&\tilde{C}^{\lambda \lambda}_\Theta(x,0,0) \approx 
\\[3mm] \nonumber
&&\approx\ln(\frac{|x|}{l}) \int_{-\frac{\pi}{a}}^{\frac{\pi}{a}} 
\frac{\rd q_y}{\frac{2\pi}{a}}
\left[\frac{\det(K_\Phi)}{K_{+} + K_{-}}
\left[(K_\Phi^{-1})^{\lambda \lambda}+
\frac{K_\Theta^{\lambda \lambda}}{K_{+}K_{-}}\right]\right] .
\eea
The charge fluctuation factor in the fermion correlation functions has a conventional
algebraic decay with finite power $d_{\Theta,x}$.  The faster than algebraic decay of the 
fermion one-particle Green's function implies that the singularity in Landau gauge
occupation numbers, a step function of unit magnitude in Hartree-Fock theory, is exceedingly weak. 

The correlation function of the $\Phi$ field along directions perpendicular to the stripes 
is given by
\bea
\nonumber
\tilde{C}^{\lambda \lambda}_\Phi(0,y,0) & \approx &
\int_{1/L}^\Lambda\frac{\rd q_x}{q_x} 
\int_{-\pi/a}^{\pi/a} \frac{\rd q_y}{\frac{2\pi}{a}}\left[1-\cos(q_y y)\right]\cdot
\\[3mm]\nonumber
&&\, \cdot\left[\frac{\det(K_\Theta)}{K_{+}+K_{-}}
\left[(K_\Theta^{-1})^{\lambda \lambda}+
\frac{K_\Phi^{\lambda \lambda}}{K_{+} K_{-}}\right]\right]
\\[3mm]
&\approx& C\ln\left(\frac{L}{l}\right)\ln\left(\frac{|y|}{a}\right) \quad ,
\eea
where $C$ is a finite constant that can be found numerically.
It follows that the corresponding factor in the one-particle Green's function 
has a faster than algebraic decay in the thermodynamic limit. 
The factor associated with charge fluctuations has a similar dependence and is 
is given by
\bea
\nonumber
{C}^{\lambda \lambda}_\Theta(0,y,0)  & \approx &
\int_{1/L}^\Lambda\frac{\rd q_x}{q_x}
\int_{-\pi/a}^{\pi/a} \frac{\rd q_y}{\frac{2\pi}{a}}\cos(q_y y) \cdot
\\[3mm]
&& \cdot \left[\frac{\det(K_\Phi)}{K_{+}+K_{-}}
\left[(K_\Phi^{-1})^{\lambda \lambda}+
\frac{K_\Theta^{\lambda \lambda}}{K_{+} K_{-}}\right]\right]
\nonumber \\[3mm]
& \approx & \ln\left(\frac{L}{l}\right) 
C_\Theta\left(\frac{a}{|y|}\right)
\quad ,
\eea
where $C_\Theta$ is a function of the ratio $(a/|y|)$ and is finite as $|y|\rightarrow +\infty$.

\subsection{Tunneling density of States }

These results for boson correlation functions may be assembled to evaluate the 
imaginary time dependence of the local fermion Matsubara Green's function and, by Laplace
transforming this, the density of states for tunneling into the bilayer system, 
a quantity that is in principle measurable. 
The single-particle Matsubara Green's function is given by
\bea
G(0,0,\tau) = \left<{R^\lambda}^\dagger_j(x,\tau) \, R_j^\lambda(x,0)\right>
\approx \exp\left\{-(1/2)\tilde{C}(\tau)\right\}, 
\eea
where
$
\tilde{C}(0,0,\tau) = 
 \tilde{C}_{\Phi}(0,0,\tau) + \tilde{C}_{\Theta}(0,0,\tau)
$
and
\bea
\tilde{C}_{\Phi/\Theta}(0,0,\tau)=2\int\frac{\rd^2 q \rd \omega}{(2\pi)^3/a}
\left(1-\cos\omega \tau \right)
\left[M^{-1}_{\Phi/\Theta}({\bf q},\omega)\right]^{\lambda \lambda} \hspace{-4mm} .
\label{matsu}
\eea
We first discuss the balanced bilayer case for which
the $K_\Phi^{\lambda \mu}$ and $K_\Theta^{\lambda \mu}$ 
matrices become simple numbers and the integral is simpler to treat analytically.
In this case $\tilde{K}_{\Phi}(q_y)$, $\tilde{K}_{\Theta}(q_y)$, are given by
(\ref{Kf-qy}-\ref{Kth-qy}) and
\bea
\tilde{C}_{\Phi}(0,0,\tau) & = & 2\int\frac{\rd^2 q \rd \omega}{(2\pi)^3/a}
\frac{\left(1-\cos\omega \tau \right)}{
\frac{\omega^2}{\pi^2\tilde{K}_{\Theta}(q_y)} + \tilde{K}_{\Phi}(q_y)q_x^2}
\nonumber\\[3mm]
&& \hspace{-21mm}
=\int \frac{\rd^2 q}{4\pi/a} \frac{1}{|q_x|}\sqrt{ 
\frac{\tilde{K}_\Theta(q_y)}{\tilde{K}_\Phi(q_y)} }
\left(1 - e^{-\pi |q_x| \sqrt{\tilde{K}_\Phi(q_y)\tilde{K}_\Theta(q_y)}\, \tau} \right) 
\, .
\label{matsub}
\eea
We can understand the content of this integral by means of the following analysis.
The integral can be separated into the sum of two terms, contributions from the 
region where $q_x$ and $q_y$ are small and the exponential can be approximated by
the first few terms of the Taylor expansion and contributions from larger $q_x$ and $q_y$ where
the exponential can be disregarded. The leading contribution to the integral comes form the 
lower boundary of the second region where $q_y\approx 1/(q_x \tau)$. 
We focus on the case of large $\tau$ for which our low energy theory applies. In 
this limit we can approximate $\tilde{K}_{\Theta}(q_y)\approx \tilde{K}_{\Theta}(0)$
and $\tilde{K}_{\Phi}(q_y)\approx \tilde{K}^{''}_{\Phi}(0)q_y^2$.  In this limit Eq.~(\ref{matsub})
becomes
\bea
\tilde{C}_{\Phi}(0,0,\tau) & = & \frac{a}{4\pi}
\sqrt{\frac{ \tilde{K}_{\Theta}(0)}{\tilde{K}^{''}_{\Phi}(0)}} \cdot
\\[3mm]\nonumber
&& \cdot \int \frac{\rd q_x \rd q_y}{|q_x||q_y|}
\left(1-e^{-\pi |q_x||q_y|\tau \sqrt{\tilde{K}_{\Theta}(0)\tilde{K}^{''}_{\Phi}(0)}}
\right)
\\[3mm] \nonumber
 & \approx & \frac{a}{2\pi} 
\sqrt{\frac{ \tilde{K}_{\Theta}(0)}{\tilde{K}^{''}_{\Phi}(0)}}
\ln^2\left(\frac{\pi^3}{al}\sqrt{\tilde{K}_{\Theta}(0)
\tilde{K}^{''}_{\Phi}(0)} \, \tau \right) \,. 
\label{tsubara1}
\eea
The Matsubara Green's function factor contributed by $\Theta$ field correlations has a weaker
$\tau$ dependence which we can neglect for the present qualitative analysis.  

Similar steps can be taken for the general case of unbalanced bilayers.  
After the $\omega$ integration, for (\ref{matsu}) we obtain
{\small
\bea
\tilde{C}_{\Phi}(0,0,\tau) & = & \frac{a}{4\pi}
\int \frac{\rd^2 q}{|q_x|} \frac{K_\Theta^{11}}{K_{+}^2-K_{-}^2}
\\[2mm]\nonumber
&& \hspace{-12mm}\left\{\left[\left(K_{+}-\frac{K_\Phi^{22}}
{\left(K_\Theta^{-1}\right)^{22}}\frac{1}{K_{+}} \right)
-\left(K_{-}-\frac{K_\Phi^{22}}{\left(K_\Theta^{-1}\right)^{22}}
\frac{1}{K_{-}}\right)
\right] \right. 
\\[2mm]\nonumber
&&-\frac{1}{K_{+}^2 - K_{-}^2} \left[
\left(K_{+}-\frac{K_{\Phi}^{22}}{\left(K_\Theta^{-1}\right)^{22}}\frac{1}{K_{+}}\right)
e^{-\omega_{+}(q_y)\tau} \right.
\\[2mm]\nonumber
&&\left. \left.
-\left(K_{-}-\frac{K_\Phi^{22}}{\left(K_\Theta^{-1}\right)^{22}}\frac{1}{K_{-}}\right)
e^{-\omega_{-}(q_y)\tau} 
\right]\right\} \quad .
\eea}
Since at small $q_y$, $K_- \sim |q_y|$, the most important contributions to the
integral will come from the terms containing $1/K_{-}$ and from the 
exponential factors containing the argument $\omega_{-}(q_y)$.  Keeping only these terms we obtain
\bea
\tilde{C}_{\Phi}(0,0,\tau) & \approx &\frac{a}{4\pi}
\frac{K_{\Theta}^{11}(0)}{K^2_+(0)K_{-}^{''}(0)}
\frac{K_\Phi^{22}(0)}{\left(K_\Theta^{-1}\right)^{22}(0)}\cdot
\\[2mm]\nonumber
&& \cdot\int \frac{\rd q_x \rd q_y}{|q_x||q_y|}
\left(1-\frac{1}{K^2_{+}(0)} e^{-\pi K_{-}^{''}(0) \tau |q_x||q_y|}\right)
\label{tsubara2}
\\[3mm] \nonumber
&\approx & \frac{a}{2\pi}
\frac{\det(K_\Theta)(0)}{K^2_+(0) K_{-}^{''}(0)}
K_\Phi^{22}(0)
\ln^2\left(\frac{\pi^3}{a l} K_{-}^{''}(0) \tau\right) \,,
\eea
demonstrating that the form of the 
Matsubara Green's function does not change qualitatively at unbalanced filling factors. 
The tunneling density of states is the inverse Laplace transform of $G(0,0,\tau)$,
\bea
G(0,0,\tau)\approx e^{-\frac{\alpha}{2} \ln^2(\Omega_0 \tau)}
=\int_0^{+\infty}\rd E\rho_{\rm tun}(E)e^{-E|\tau|} \quad ,
\eea
where $\alpha$ and $\Omega_0$ can be identified from
Eq.~(\ref{tsubara1}) for balanced bilayers and from Eq.~(\ref{tsubara2}) in the unbalanced case. 
The inverse Laplace transform of such a function is
\bea
&&\rho_{\rm tun}(E) = 
\\[3mm]\nonumber
&&= 
\frac{e^{-(\alpha/2)\ln^2(\Omega_0/E)}}{E/\Omega_0} \sum_{k=0}^\infty
 \frac{\tilde\Gamma_k(1)}{k!} (-\frac{\sqrt{\alpha}}{\sqrt{2}})^{k+1}H_{k+1}
\left[\frac{\sqrt{\alpha}}{\sqrt{2}}\ln\frac{ E}{\Omega_0}\right]  ,
\eea
where $\tilde\Gamma_k(1)$ is the $k$th derivative of $1/\Gamma(r)$
at $r=1$ and $H_k$ are the Hermite polynomials. In the asymptotic case of small
$E$ (low energies) we have
\bea
&&\rho_{\rm tun}(E)\sim \frac{e^{-(\alpha/2)\ln^2(\Omega_0/E)}}{E/\Omega_0}
\frac{\alpha \ln(\Omega_0/E)}{\Gamma\left(1+\alpha \ln(\Omega_0/E)\right)}
\\[3mm]\nonumber
&& \approx  \exp\left\{-\frac{\alpha}{2}\ln^2\frac{\Omega_0}{E} - 
(\alpha \ln\frac{\Omega_0}{E} -\frac{1}{2})
\ln \ln \frac{\Omega_0}{E} + \ln \frac{\Omega_0}{E} 
\right\}\, .
\eea
This shows that the density of states vanishes at the fermi energy stronger than any 
power of $E$ and that in this sense the stripe edge physics in bilayer quantum Hall systems 
is not that of a usual system of weakly
coupled Luttinger liquids.  The above result generalizes to the case of bilayer systems, points
that have been made about single-layer quantum Hall stripe states by 
Lopatnikova {\em et al}\cite{Lopatnikova} and Barci {\em et al}\cite{Barci}. 

\section{Stability of the quantum hall smectic phase}
\label{stability}


We now consider ``backward'' (interchannel) scattering interactions that do not
conserve the number of electrons in each stripe edge.  The most important conclusion of 
the following analysis is that back-scattering interactions are much more important for  
bilayer stripe states than for single-layer stripe states. 
We can classify the back-scattering interactions as either intralayer interactions that involve
electrons only in one layer or interlayer interactions that involve electrons in both layers.
Non-zero back-scattering two-particle matrix elements conserve total momentum along the stripes, which means
that the two Landau gauge guiding center jumps must sum to zero.  In the case of quantum Hall
stripe states, the microcopic matrix elements associated with these back-scattering processes
tend to be small and these interactions will be important
only if they produce strong infrared divergences in perturbation theory.  The strength of 
these divergences is characterized here by evaluating lowest order perturbative renormalization group
scaling dimensions for these operators. 
As in Ref.~\onlinecite{MacDonald} our renormalization group (RG) scheme involves only 
$x$ and $\tau$ dimensions and and treats the rung label as an internal index of the fields 
$\Phi^\lambda_j$, $\Theta^\lambda_j$.  The philosophy underlying this procedure 
is discussed in the next section. 

\subsection{Interlayer Tunneling} 

We first address the tunneling of an electron from one stripe edge
to the other in the same rung. The action contribution from this
process has the following bosonized form: 
\begin{equation}
{\cal S}_{\rm Tunn}=-u\int \rd x \rd \tau\sum_{j}\sum_{\lambda}
\left[\exp(i2\Theta_{j}^{\lambda})+{\rm h.c.}\right] \quad ,
\label{actiontunnel}
\end{equation}
where the microscopic amplitude $u$ is discussed below. We integrate out ``fast" boson
modes $\Phi^{\lambda}$, $\Theta^{\lambda}$
in a shell, with $\Lambda/b < |q_{x}| < \Lambda$ and $\omega,
q_{y}$ unrestricted, and then rescale $q_{x}^{\prime} = b q_{x}$ and
$\omega^{\prime} = b \omega$ leaving $q_{y}$ unchanged. With an appropriate
rescaling of $\Phi$, this RG transformation leaves the harmonic smectic action
$S_{0}$ [and the dispersion relation (\ref{modes0}) for $K_{\Phi/\Theta}$ 
(\ref{HFrealspaceK12})], invariant. Stability of the smectic fixed point in the 
presence of back-scattering can be tested by considering the lowest order RG flow 
equation,
\begin{equation}
\frac{\partial u}{\partial t}=(2-\Delta_{{\rm Tunn}})u \quad ,
\end{equation}
with $t=\ln b$. As can be seen from this equation the tunneling operator will 
become relevant when its scaling dimension is less than two. 
When this term in the Hamiltonian is strong, the system is described at low energies by 
a set of quantum sine-Gordon models coupled by gradient terms (\ref{thetaaction}). 
In the domain $0< \Delta_{\rm Tunn} <2$ the continuous symmetry, present in (\ref{thetaaction}),
and broken by tunneling, is lost in the low-energy fixed point action.  This model has a discrete
symmetry $\Theta_j^\lambda \rightarrow \Theta_j^\lambda +\pi n$ for any integer
$n$ and the QFT becomes massive\cite{papa}.  The gap due to tunneling will lead to an integer quantum Hall
effect at total filling factor $\nu_T=1$. 
Using Eqs.~(\ref{thetaaction}), and (\ref{actiontunnel}),
we find the following expression for the scaling dimension
\begin{eqnarray}
\Delta_{\rm Tunn} & = &
\int_{-\pi/a}^{\pi/a}\frac{\rd q_{y}}{\frac{2\pi}{a}} 
\Biggl[\frac{\det K_{\Phi}}{K_{+} + K_{-}}.
\nonumber\\[3mm]
& & \qquad\qquad\cdot\Biggl(\left(K_{\Phi}^{-1}\right)^{11}
+\frac{K_{\Theta}^{11}}{K_{+}K_{-}}\Biggr)\Biggl]
\quad . 
\label{dimtunnel}
\end{eqnarray}
The integrand in the integral over $q_{y}$ is similar to that involved in the $\Theta$ 
boson correlation field and, ignoring the matrix character of the coefficients that appear 
in the smectic fixed point Hamiltonian, is $\sim \sqrt{K_{\Phi}/K_{\Theta}}$.  Since 
$K_{\Phi}$ vanishes for $q_y \to 0$, we can expect this quantity to be small.  Indeed we 
find by evaluating this integral numerically that interlayer tunneling is always relevant.  

\subsection{Coulomb Backscattering Interactions} 

The tunneling amplitude in bilayer quantum Hall systems can be made extremely small by making the barrier
between quantum wells higher or wider and is often completely negligible in practice.
Coulomb interactions,
on the other hand, are always present and must always be considered.  
We consider {\em interlayer} and {\em intralayer} Coulomb back-scattering processes separately.  
In the strongest interlayer back-scattering process
an electron is transferred from, say, a left-moving top layer stripe edge 
to a right-moving edge in the same rung pair of the same layer, while in the same rung pair
of the bottom layer an electron is transferred in the opposite direction,
as depicted in Fig.~\ref{fig1}.
The interlayer back-scattering operators for processes involving neighboring rungs have 
large scaling dimensions and tend to be irrelevant. 
In addition, the bare matrix elements for such a process will fall off
rapidly in magnitude with increasing distance between the rungs involved.
The action for interlayer back-scattering interactions for electrons 
within the same rung pair reads
\begin{equation}
{\cal S}_{\rm inter}=-u\int \rd x \rd \tau\sum_{j}
\left[\exp(i2(\Theta_{j}^{1}+\Theta_{j}^{2}))+{\rm h.c.}\right]
\quad .
\label{actioninter}
\end{equation}
[The other kind of process involving two neighboring rung pairs
is related to the above one by a particle-hole transformation
$\nu\mapsto 1-\nu$]. 

After an elementary calculation we obtain the following scaling dimension expression: 
\begin{eqnarray}
\Delta_{\rm inter} & = &
\int_{-\pi/a}^{\pi/a}\frac{\rd q_{y}}{\frac{2\pi}{a}} 
\Biggl[\frac{2\det K_{\Phi}}{K_{+} + K_{-}}
\nonumber\\[3mm]
& & \cdot\Biggl(
\left(K_{\Phi}^{-1}\right)^{11}-\left(K_{\Phi}^{-1}\right)^{12}
+\frac{K_{\Theta}^{11}-K_{\Theta}^{12}}
{K_{+} K_{-}}\Biggr)\Biggl]\quad .
\label{diminter}
\end{eqnarray}
This expression is similar to that which would be obtained for inter-wire back-scattering
interactions in a systems of two coupled quantum wires.  This integral is similar to the one that
appears in the tunneling operator scaling dimension calculation, although it is easy to
verify that forward scattering interactions between different stripes play an essential role.
As we discuss below, this operator is usually strongly relevant ($\Delta_{\rm inter}\rightarrow 0$),
so that at low temperatures the phases $\Theta^1_j$ and $\Theta^2_{j}$ of neighboring 
two edge system are strongly anti correlated. 
The low energy nontopological (chargeless) excitations in this limit can be understood by approximating 
$\cos[(\Theta_j^1+\Theta_j^2)] \approx 1-(\Theta_j^1+\Theta_j^2)^2/2$. 
When a term of this form is added to the quadratic Hamiltonian, the low-energy collective mode
dispersion at long-wavelengths takes the form of a spatially anisotropic two-dimensional XY
ferromagnet, $\omega^2 \sim K q_x^2 + u q_y^2$.  We discuss the significance of this result at 
greater length below. 

Finally, in an intra-layer back-scattering process two electrons move
in opposite directions between pairs of stripe edges in the same layer 
with the same separation (cf. Fig~\ref{fig1}).
Here we also concentrate on processes involving
neighboring rungs only. Processes involving two rung pairs are again related
to those involving three pairs by a particle-hole transformation.
For the first case the action reads
\begin{eqnarray}
{\cal S}_{\rm intra} & = & -u\int {\rm d} x {\rm d}\tau\sum_{j}
[\exp(i(\Phi^{2}_{j}-\Phi^{1}_{j}
+\Phi^{1}_{j+1}-\Phi^{2}_{j+1}))\nonumber\\
& & \cdot\exp(i(-\Theta^{2}_{j}-\Theta^{1}_{j}
+\Theta^{1}_{j+1}+\Theta^{2}_{j+1}))
+{\rm h.c.}] \, ,
\label{actionintra}
\end{eqnarray}
which leads to a scaling dimension $\Delta_{\rm intra}=\Delta_{\Phi}+\Delta_{\Theta}$
with
\begin{eqnarray}
\Delta_{\Phi} & = &
\int_{-\pi/a}^{\pi/a}\frac{\rd q_{y}}{\frac{2\pi}{a}} 
\Biggl[\frac{\det K_{\Theta}}{K_{+} + K_{-}}
\left(1-\cos(q_{y}a)\right)\nonumber\\[3mm]
& & \cdot\left(\left(K_{\Theta}^{-1}\right)^{11}
+\left(K_{\Theta}^{-1}\right)^{12}\
+\frac{K_{\Phi}^{11}+K_{\Phi}^{12}}
{K_{+} K_{-}}\right)\quad ,
\label{dimphi}
\\[3mm]
\Delta_{\Theta} & = &
\int_{-\pi/a}^{\pi/a}\frac{\rd q_{y}}{\frac{2\pi}{a}} 
\Biggl[\frac{\det K_{\Phi}}{K_{+} + K_{-}}
\left(1-\cos(q_{y}a)\right)
\nonumber\\[3mm]
& & \cdot\left(\left(K_{\Phi}^{-1}\right)^{11}
-\left(K_{\Phi}^{-1}\right)^{12}\
+\frac{K_{\Theta}^{11}-K_{\Theta}^{12}}
{K_{+} K_{-}}\right)\quad .
\label{dimtheta}
\end{eqnarray}
Note that the imaginary part of the integrand in Eqs.~(\ref{diminter}),
(\ref{dimphi}-\ref{dimtheta}) does not contribute to the integrals.
Backscattering processes other than the ones discussed above have larger scaling 
dimensions and also involve
larger momentum transfer and have therefore exponentially smaller
bare matrix elements. We therefore shall concentrate on the
processes discussed above.


We now discuss our numerical results for scaling dimensions of the operators that 
are not described by the quadratic boson theory.  
Fig.~\ref{fig2} shows the scaling dimensions of the back-scattering 
interactions in a balanced system ($\nu=1/2$) in the second excited
Landau level ($N=2$) as a function of the layer separation $d/\ell$.
The stripe period chosen for these calculations, 
$a=5.8\ell$,  corresponds to the period at which the 
Hartree-Fock energy of the stripe state in a isolated layer is minimized.~\cite{Jungwirth}.
Interestingly, single-electron tunneling is irrelevant ($\Delta_{\rm Tunn} >2$) for
$d/\ell > 1.5$, but is strongly relevant at smaller layer separations.
Interlayer back-scattering is relevant
($\Delta_{\rm inter}<2$) at all layer separations, more strongly so at 
smaller layer separations, while the scaling
dimension of the intra layer back-scattering is smaller than 2 only for
$d/\ell\gtrsim 2$ and approaches a value of $\Delta_{\rm intra}\approx 1.84$
for $d/\ell\gg 1$.  This value for the limit of weak interactions between
the layers recovers the single-layer result obtained earlier.~\cite{MacDonald}.
The contributions $\Delta_{\Phi}$ and $\Delta_{\Theta}$ 
to $\Delta_{\rm intra}$, not shown in the figure, become equal in this case. 
%
%
%
%
\begin{figure}
\centerline{\includegraphics[width=8cm]{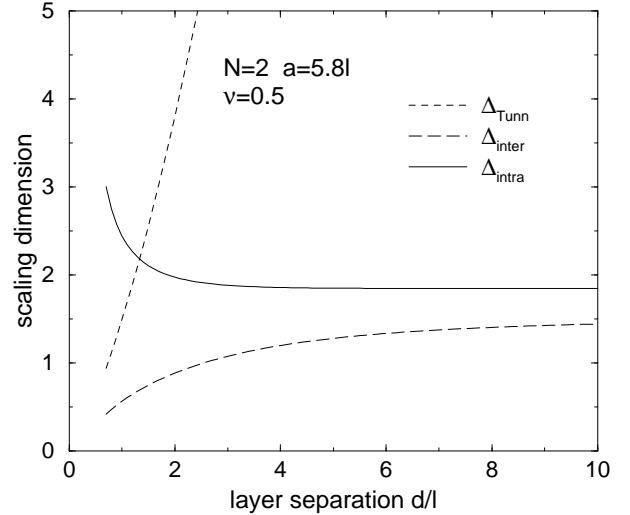}}
\caption{Scaling dimensions for tunneling and back scattering interactions
as a function of layer separation $d/l$ in a balanced bilayer system at bilayer 
total filling factor $\nu_T=9$,
{\it i.e.} with a $N=2$ valence Landau level. 
\label{fig2}}
\end{figure}
\begin{figure}
\unitlength=1mm
\begin{picture}(80,31)
\put(-3.8,3){\epsfig{file=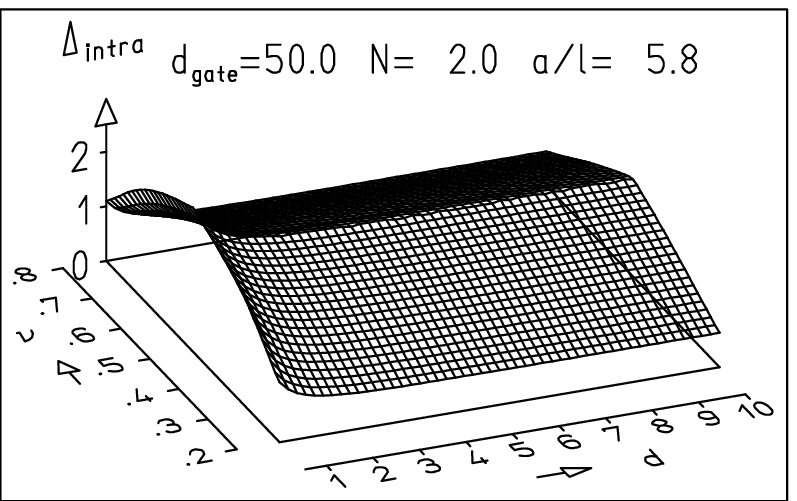,height=27mm}}
\put(39.0,3){\epsfig{file=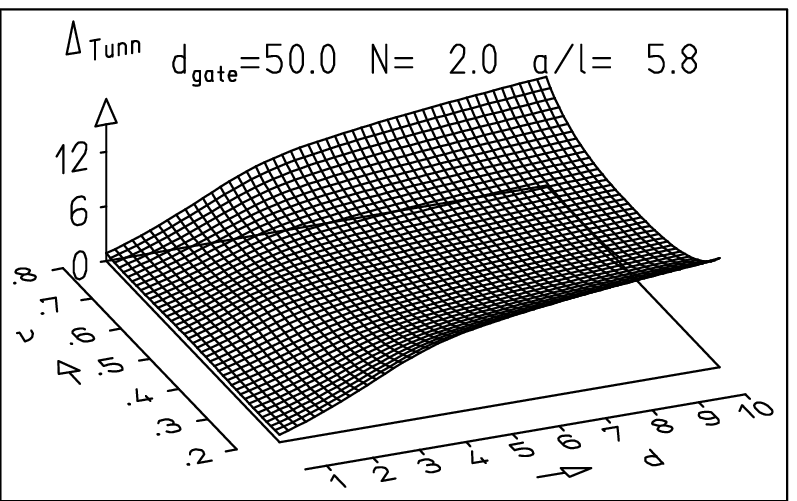,height=27mm}}
\end{picture}
\caption{The left panel shows the $N=2$, $d/\ell=5.8$, intralayer back-scattering interaction
scaling dimension. The right panel shows the scaling dimension
of the Tunneling operator.}
\label{fig3}
\end{figure}
\begin{figure}
\unitlength=1mm
\begin{picture}(80,31)
\put(-3.8,3){\epsfig{file=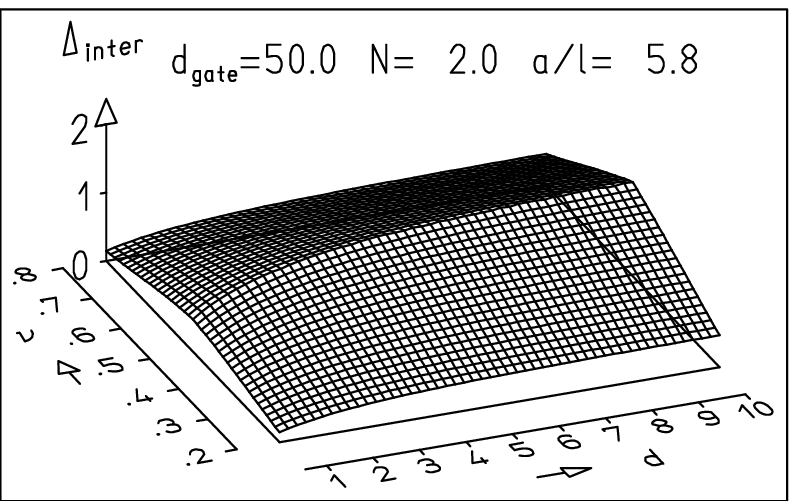,height=27mm}}
\put(39,3){\epsfig{file=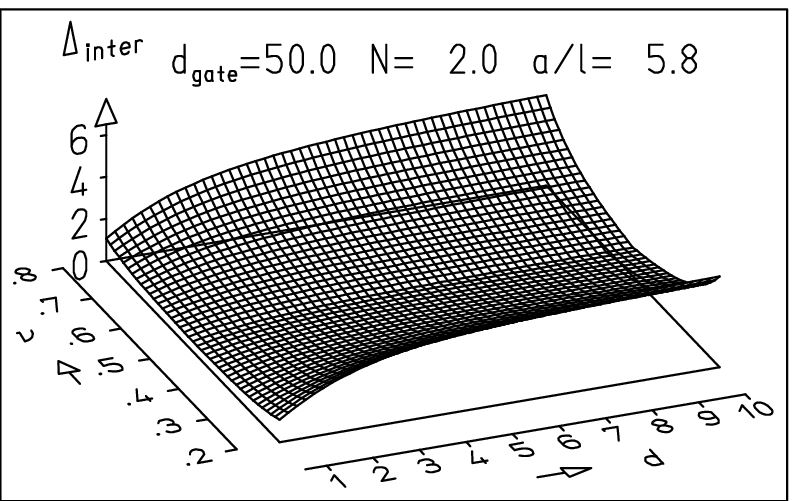,height=27mm}}
\end{picture}
\caption{The left panel show the scaling dimensions for interlayer back-scattering across
narrow and the right panel for interlayer back-scattering across wide rungs in
unbalanced bilayers.  Note the difference in scale between left and right panels.
The most relevant interlayer back-scattering interactions are those of narrow rungs.
}
\label{fig4}
\end{figure}

The dependence of scaling dimensions on the bilayer balance is illustrated in 
Fig.~\ref{fig3} and Fig.~\ref{fig4}.  We see that intralayer interactions become more relevant when
their individual filling factors move away from $\nu=0.5$, as in the single-layer 
case, while the tunneling operator becomes less relevant.   
Interestingly the interlayer back-scattering interactions show different results
depending on the distance between the edges involved in the transition.
For $\nu\neq 0.5$ we have to distinguish between nearest neighbor
interlayer and intralayer back-scattering processes that involve, 
according to the definition given in Fig.~\ref{fig1}, 
only the smallest number of neighboring rung pairs
(one and two, respectively) and those processes that involve formally
two and three rung pairs, respectively. These two kinds of processes are
related by particle-hole transformation and therefore shown in 
different panels.  Generally the scaling dimension increases with the distance between
the edges. The data shows that one of these two back-scattering processes, related
by a particle-hole transformation, is always relevant and that the minimum 
scaling dimension decreasing with increasing bias between the layers.
In summary, the most relevant residual interactions is interlayer back-scattering
and they are increasingly important as the bilayer is unbalanced. 

%
%

We note that the scaling dimensions $\Delta_{\rm Tunn} $ of the single-electron 
tunneling and $\Delta_{\rm inter}$ of the interlayer back-scattering
approach zero for $d/\ell\to 0$, i.e. these processes become strongly
relevant. This is a natural result since in this limit we recover
the monolayer electron spin quantum Hall ferromagnet. This system is
perfectly 
\begin{figure}
\centerline{\includegraphics[width=8cm]{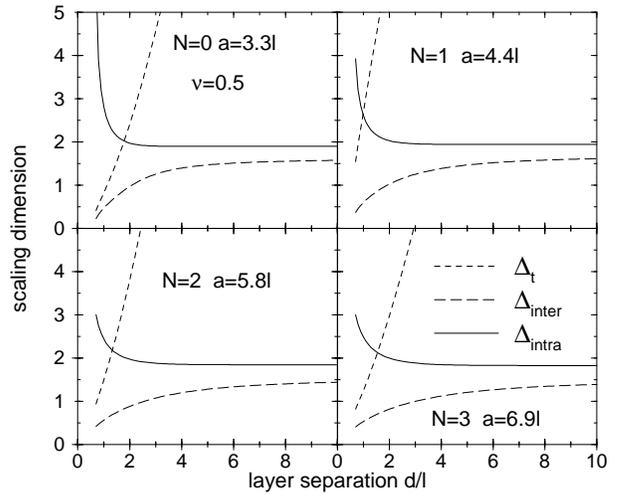}}
\caption{The scaling dimension of the back-scattering interactions
in the various Landau level ($N=0,1,2,3$) in a balanced system as a
function of the layer separation $d/\ell$. The stripe periods $a$  are obtained from
Hartree-Fock monolayer results given in Ref.~\protect{\cite{Jungwirth}}.
\label{fig6}}
\end{figure}
\hspace{-3.2mm}isotropic in pseudospin space, and therefore processes like
tunneling which acts essentially like a (pseudo-)magnetic field are
obviously very relevant.  This increased relevance arises formally 
in our calculations through the property that the matrices
$K_{\Phi}(q_{y})$ vanish in the limit $d \to 0$, so that the integrands in 
Eq.~(\ref{dimtunnel}) and (\ref{diminter}) are identically zero.  $K_{\Phi}$ vanishes
because it is a measure of energy changes associated with charge transfer between
layers at a particular stripe edge; when $d \to 0$ only the total charge near
each stripe edge influences the energy functional.
%
%
In Fig.~\ref{fig6} we show the dependence of these scaling dimensions on 
Landau level index $N=0,1,2,3$ with the 
stripe periods taken from Ref.~\cite{Jungwirth} in each case. As this figure shows,
our results for the scaling dimensions of back-scattering processes
around the assumed stripe state depend only weakly on the Landau level
index. We note that in the lowest and first excited Landau level ($N=0,1$)
no conductance anisotropies are found experimentally in single layers, even though
there is a stripe state in each of these Landau levels
which is a local minimum of the Hartree-Fock energy functional.  The true ground
state in these instances is far from the stripe state, differencing in character 
even at microscopic length scales.  The fact that our calculation does not 
obtain anomalous results in cases where we do not believe stripe states occur, emphasizes again
that our approach can only address the properties of systems in which fluctuations around
Hartree-Fock stripe states are weak.  It cannot predict when stripe states occur.
Future experimental activity will be necessary to identify with confidence when stripe
states occur in bilayers. 


\subsection{Smectic interlayer phase coherent and Wigner Crystal states}

\subsubsection{Smectic interlayer phase coherent state}

In this section we examine the effect of the interlayer back scattering interactions
(when they are strongly relevant) on the low energy physics of the system
and show that the phase coherence is marked by a nonvanishing value of an
interlayer phase order parameter.  In this phase electrons at each stripe edge are
coherent superpositions of the upper and lower layer states.

The most relevant interlayer back-scattering operators are related by particle-hole
symmetry and describe back-scattering across an electron stripe in the top layer and
the corresponding hole stripe in the bottom layer or across a hole stripe in the 
top layer and the corresponding electron stripe in the bottom layer.  In terms of 
the Luttinger liquid fields we have defined, the sum of these two interactions takes the 
form: 
\bea
\hat{O}_{\rm inter}&=&-u_{\rm inter,\nu} \cos[2(\Theta_j^1 +\Theta_j^2)]
\\[3mm] \nonumber
&& \hspace{10mm}
- u_{\rm inter,(1-\nu)} \cos[2(\Theta_j^2 +\Theta_{\rm j+1}^1)] \quad.
\eea
Expressions for the bare values of these coupling constants are given below.
As shown on Figs.~5(a), 5(b), (left, right panel respectively), at small
layer separations these operators are strongly relevant. At low temperatures 
the phases $\Theta^1_j$ and $\Theta^2_{j}$, $\Theta^2_j$ and $\Theta^1_{j+1}$ of neighboring
two edges tend to be strongly  anti correlated.
The low energy excitations in this limit can be understood by approximating
$\cos[(\Theta_j^1+\Theta_j^2)] \approx 1-(\Theta_j^1+\Theta_j^2)^2/2$. When terms of this
form are added to the action, it takes the following form:
\begin{eqnarray}
{\cal S}_{\Theta} & = & \frac{1}{2}\int_{{\bf q},\omega}
\sum_{\lambda,\mu}\biggl[\Theta^{\lambda *}({\bf q},\omega) {\cal M}^{\lambda \mu}_\Theta
\, \Theta^{\mu } ({\bf q},\omega) \quad.
\eea
The new matrix ${\cal M}_\Theta$ is given by
\bea
{\cal M}_\Theta = M_\Theta + 
2\left( \begin{array}{cc}
u_{{\rm i}1}+u_{{\rm i}2} \;& u_{{\rm i}1} + u_{{\rm i}2} e^{-iq_y a}\\ \\
u_{{\rm i}1} + u_{{\rm i}2} e^{iq_y a} \; & u_{{\rm i}1}+u_{{\rm i}2} 
\end{array}\right) \, ,
\label{N_Theta}
\eea
where $M_\Theta$ is the matrix of the system at the smectic fixed point and is given by 
Eq.~(\ref{thetaaction}) or by Eq.~(\ref{M_matrix}) (interchanging $K_\Phi$, $K_\Theta$) and
$u_{{\rm i}1}$, $u_{{\rm i}2}$, is the short notation for 
$u_{\rm inter,\nu}$, $u_{\rm inter,(1-\nu)}$, 
respectively.
The effects of the interlayer back-scattering interactions, included on the new matrix of 
Eq.~(\ref{N_Theta}) (which we denote by $N_\Theta$), shift the poles of the boson propagators.
The low energy collective modes now are given by
\bea
\omega^2_\pm({\bf q})=\pi^2\frac{A}{2}\left[1 \pm \sqrt{1 -\frac{4D}{A^2}} \right] \quad ,
\eea
where
\begin{figure}
\unitlength=1mm
\begin{picture}(80,29)
\put(-1.5,3){\epsfig{file=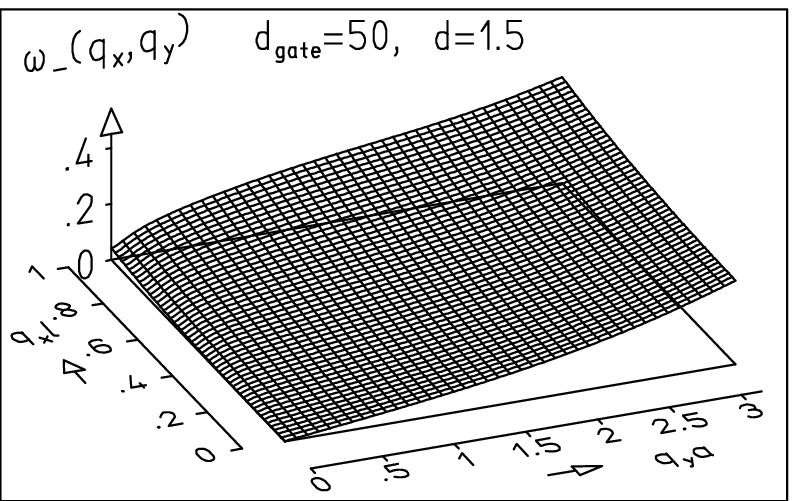,height=25mm}}
\put(38.5,3){\epsfig{file=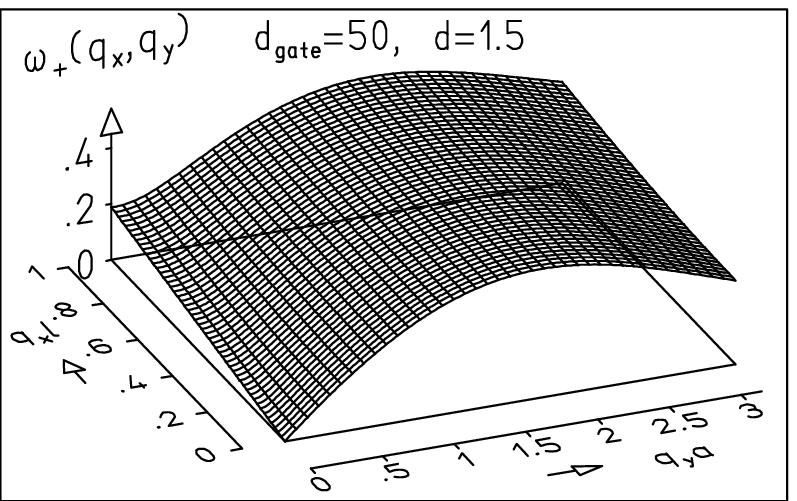,height=25mm}}
\end{picture}
\caption{The collective modes of the bilayer QH smectic interlayer coherent
 phase, for the case of
$K_{\Phi/\Theta}(x,j)$ of Eqs.~(\ref{HFrealspaceK12}-\ref{W}) and $N=2$, $\nu\approx1/2$ and $a=5.8 l$,
are shown.
$\omega_{+}(q_x,q_y)$, the right panel, is of the form of a spatially anisotropic two-dimensional
ferromagnet, $\omega_{+}^2({\bf q}) \sim K q_x^2 + u q_y^2$.
The $\omega_{-}(q_x,q_y)$ collective mode vanishes only for $q_y \to 0$ {\em and} $q_x \to 0$
when non-local contributions to the interaction coefficients are accounted for.
}
\label{modes2}
\end{figure}
\bea
A=q_x^2 {\rm Tr}(K_\Phi K_\Theta) + 2N_\Theta^{11} K_\Phi^{22} + 
2 \Re\left\{ K_\Phi^{12} N_\Theta^{21}\right\} \, ,
\eea
\bea
D &=& \det(K_\Phi) \left\{q_x^4 \det(K_\Theta) + 2N_\Theta^{22} K_\theta^{11} +
\det(N_\Theta)
\right.
\\[3mm]\nonumber
&&\hspace{15mm}\left.
 - 2 q_x^2 \Re\left\{K_\Theta^{12} N_\Theta^{21}\right\}
\right\} \quad .
\eea
These low energy collective modes are shown in Fig.~\ref{modes2}.


As before, the case of balanced filling fraction can be described in a more transparent way 
using the scalar couplings ${\tilde K}_\Phi$, ${\tilde K}_\Theta$ of Eq.~(\ref{Kf-qy}),
(\ref{Kth-qy}).
For this case the low-energy collective modes have the form 
\bea
&&\omega^2_\pm({\bf q})=
\nonumber
\\[3mm]
&&=\pi^2 {\tilde K}_\Phi({\bf q})\left[q_x^2 {\tilde K}_\Theta ({\bf q})
+ 2 u_{{\rm i}1} \left(1+\cos(q_y a/2)\right)
\right]\quad .
\eea
In this formulation one of the gapless modes is located at the edge
of the Brillouin zone, which is now
doubled, at $q_y=2\pi/a$.
In the extended Brillouin zone we use for balanced bilayers the $\omega_-({\bf q})$
softmode appears for $q_y\rightarrow 0$, whereas the $\omega_+({\bf q})$
softmode appears as $q_y\rightarrow 2\pi/a$.  The two modes have the following behaviors:
\bea
\omega_-^2({\bf q}) & \approx & \pi^2 {\tilde K}^{''}_\Phi({\bf 0}) q_y^2 
\left[{\tilde K}_\Theta ({\bf 0})q_x^2 + 4 u_{{\rm i}1} \right] \quad,
\\[3mm]
\omega_+^2({\bf q})& \approx & \pi^2 {\tilde K}_\Phi(\frac{2\pi}{a}) 
	\left[{\tilde K}_\Theta (\frac{2\pi}{a})q_x^2 + \frac{u_{{\rm i}1} a^2}{4} 
\left(q_y - \frac{2\pi}{a}\right)^2
\right].
\eea
Similar results can be obtained using the matrix formulation for general $\nu$ and 
become equivalent for $\nu=1/2$.  There is no qualitative change in the collective 
mode structure when $\nu \ne 1/2$.

The interlayer phase coherent smectic state is characterized by a finite value 
of the following order parameter,
\bea
{\Psi}({\bf r}) &=& \left<\psi_T^\dagger({\bf r}) \psi_B({\bf r})\right>
 =\frac{1}{2\pi} \left<e^{- 2 i\Theta({\bf r})} \right>
\approx \frac{1}{2\pi} e^{-\frac{1}{2}\left<4\Theta^2({\bf r})\right>}
,
\eea
where $\psi_T^\dagger$, $\psi_B$ are fermion creation and annihilation for the
top and bottom layers respectively.
We now show that $\left< \Theta^2({\bf r})\right>$ is finite.
We discuss only the case of balanced bilayers, using the alternative formulation
which is more transparent.  We find that 
\bea
&& \left<{\Theta^2_j}\right> = 
\nonumber\\[3mm] \nonumber
&& \int \frac{d^2 q}{\frac{(2\pi)^2}{a/2}} \frac{d\omega}{2\pi} \frac{\tilde{K}_\Phi({\bf q})}{
\frac{\omega^2}{\pi^2} + \tilde{K}_\Phi({\bf q}) \left[q_x^2 \tilde{K}_\Theta({\bf q}) + 2 u_{{\rm i}1}\left[1+
\cos(\frac{q_y a}{2})\right] \right] 
}
\\[3mm]
&& \approx \frac{a}{16\pi}\int_{-2\pi/a}^{2\pi/a}\rd q_y \sqrt{\frac{\tilde{K}_\Phi(q_y)}{\tilde{K}_\Theta(q_y)}} 
\ln\left[\frac{2 \tilde{K}_\Theta(q_y) (1/l)}{u_{{\rm i}1}(1+\cos(q_ya/2))}\right]
\quad .
\label{order_par}
\eea
In (\ref{order_par}) we have introduced an upper short distance cut-off $1/\ell$
for the $q_x$ integration.
Interlayer back scattering interactions have cut-off the infrared divergence of the
$q_x$ integration, making the integral finite and establishing particle-hole pair condensation.
In this state the U(1) symmetry associated with conservation of total 
charge difference $N_T -N_B$ between top and bottom layers is broken. 

We conclude on the basis of this analysis that interlayer back-scattering will drive the 
Hartree-Fock bilayer smectic state to a state which has both broken translational and 
orientational symmetry {\em and} spontaneous interlayer phase coherence along the edges.  We expect 
this state to exhibit giant interlayer tunneling conductance anomalies at low-bias voltages,
similar to those that have been seen in the $N=0$ Landau level in bilayers.  Although 
these states have a charge gap that we discuss below and should exhibit the quantum
Hall effect, we expect that they will exhibit strongly anisotropic dissipative transport 
at finite temperatures.  Their two gapless collective modes arise because they have 
broken translational and orientational symmetry and spontaneous interlayer phase coherence.
We also note that the quantum character of 
these bilayer smectic states is quite distinct from the quantum smectics discussed
previously for the single-layer case.  For instance the long-wavelength behavior of the 
quantized collective mode $\omega_{-}(q_x,q_y)$ changes from being proportional to 
$|q_x q_y|$ to being proportional to $|q_y|$ only when spontaneous inter-layer phase
coherence is present; locking the phase difference between different layers qualitatively
increases the cost of independent position fluctuations.  The long-wavelength behavior
of the $\omega_{+}(q_x,q_y)$ collective mode is that of an anisotropic superfluid.
As in the case of uniform states, spontaneous interlayer phase coherence is equivalent to 
electron-hole pair superfluidity, but the broken orientational symmetry of the smectic
state causes this superfluid to have orientation dependent stiffness.

It seems quite possible that the order parameter that 
characterizes the broken orientational and translational symmetry of these
states will be driven to zero when interlayer interactions are sufficiently strong.
Indeed this is suggested\cite{Brey,Cote} by mean-field calculations.
We are unable to estimate where this transition takes place using the methods
of this paper. 


\subsubsection{Coherent smectic state specific heat}

The internal energy of the bilayer smectic phase coherent (SPC) state will be dominated at low 
energies by the contribution from the $\omega_{-}({\bf q})$ mode.
The leading contribution to the integral for the internal energy comes from the region 
of small ${\bf q}$.  The $q_x$ integral now has a natural infrared cut-off however 
at $\sqrt{u_{{\rm i}1}/{\tilde K}_\Theta ({\bf 0})}$.
It follows that the internal energy is given for small $u$ by 
\bea
U \approx 
\frac{2a}{\pi^3 \sqrt{\tilde{K_\theta}(0) \tilde{K}_\Phi^{''}(0)} }
T^2 \zeta(2)\ln\left(\frac{{\tilde K}_\theta(0)}{4u_{{\rm i}1}} \frac{1}{al}\right) \quad ,
\eea
and the specific heat will now be linearly dependent on $T$.  The
specific heat anomaly noted previously for the bilayer smectic is suppressed when
inter-layer coherence is established, even though broken translational and orientational
symmetry are still present. 

\subsubsection{Wigner crystal state}

Intralayer back-scattering interactions take the form 
\begin{eqnarray}
{\hat O}_{\rm intra} & = & -u
[\exp(i(2k_Fx+\Phi^{2}_{j}-\Phi^{1}_{j}
+\Phi^{1}_{j+1}-\Phi^{2}_{j+1}))\nonumber\\
& & \cdot\exp(i(-\Theta^{2}_{j}-\Theta^{1}_{j}
+\Theta^{1}_{j+1}+\Theta^{2}_{j+1}))
+{\rm h.c.}] \quad ,
\label{actionintra}
\end{eqnarray}
where the oscillatory dependence on coordinate along the edge
which we have exhibited explicitly follows from our earlier field operator 
definitions.  This interaction dominates only at quite large layer 
separations.  When it does it drives the system to a state which has periodicity
along the stripe edges as well as across the stripes.  
Since, the wavelength along the stripe is $4\pi l^2/a$, and since the periodicity 
along the direction perpendicular to the stripes is $a$, this state will contain one electron per layer
per two dimensional unit cell.  We therefore identify this state as a bilayer Wigner crystal (WC)
state. 

\subsection{Gap Estimates for Bilayer Stripe States} 

The most important conclusion from the above calculations is that interlayer Coulomb 
back-scattering interactions are always relevant in bilayer stripe states.  {\em 
The gapless bilayer stripe state can never be the true ground state.}  Since the bare 
matrix elements associated with these interactions are often quite small, however,
they will often be important only at low temperatures.  As we explain below,
we believe that Coulomb interactions
will most often drive the system either to an isotropic coherent state or to a 
smectic coherent state.  Both states will have a charge gap and an integer 
quantum Hall effect.  In this section we estimate the size of this gap and hence the 
temperature above which we expect the phenomenology of these states to cross 
over from quantum Hall behavior to stripe state behavior.

Our estimates are built on bare matrix elements whose evaluation we discuss below and 
on the scaling dimension calculations discussed above.  Given dimensionless inter- and intralayer
back-scattering interactions $u_{\rm inter}$ and $u_{\rm intra}$, we can estimate the gap
by integrating the RG flow equations to obtain
\bea
E_g^{e/a}(u_{e/a}) = b^{-1}E_g^{e/a}(b^{2-\Delta_{e/a}} u_{e/a} ) \quad ,
\eea
where the super- and subscripts $e/a$ are used for int{\it e}r- and intr{\it a}layer
interactions respectively.
When the interactions become of order 1 on the renormalized energy scale
($b^{2-\Delta} u=1$), the energy gap should be roughly
equal to the renormalized characteristic Coulomb energy $E_c$, giving
\bea
E_g(u) = (U/E_c)^{1/(2-\Delta)} E_c \quad ,
\eea
where $U=u E_c$ is the microscopic high-energy-scale back-scattering interaction strength.
The $\nu$ dependence of the gap enters through
$U$, and through the scaling dimensions.
Both effects conspire to strongly reduce the gap magnitude near half filling. Taking
$E_c=0.3e^2/l$, approximately the maximum correlation energy per electron in a partially filled
Landau level, the resulting gaps for $N=2$ and $d_{\rm gate}= 50 l$, are shown 
as a function of filling fraction
and distance between layers in Fig.~{\ref{gap_inter} and Fig.~{\ref{gap_intra}.
We notice that the gap resulting from the intralayer back-scattering interaction is very small near 
half filling, dropping below the range accessible to dilution fridges over most of 
the filling factor shown in this figure.
On the other hand the gap resulting from the interlayer back-scattering interactions is not
as small and remains reasonably large out to large values of the interlayer separation $d$.
Recalling that this interaction is proportional to $ R^{\dagger}_{1} L_{1} L^{\dagger}_{2} R_{2}$, 
we see that when this interaction is strong it favors interlayer phase coherence along 
each stripe edge and that when it is very strong it leads to condensation of the field
$\Theta_j^{1}+\Theta_j^{2}$ to a value independent of $j$.  Since $\Theta_j^{\lambda}$ is
by definition the phase difference between left and right-going fermion fields at 
the $(j,\lambda)$ stripe edge, and since the layers indices of right and left going fermions 
is opposite at $\lambda=1$ and $\lambda=2$ stripe edges, what is condensing when this interaction
is strong is the phase difference between fermions in opposite layers.  In other words, {\em the
state that occurs in the strong interedge back-scattering limit has spontaneous interlayer phase 
coherence}.  States with interlayer phase coherence and stripe order can occur as local and 
even global minima of Hartree-Fock energy functionals.  Coupled with the irrelevance of intralayer
back scattering interactions at small layer separations in the bilayer case, our analysis
suggests that they can be the ground states of bilayer quantum Hall systems in high Landau levels.

\begin{figure}
\unitlength=1mm
\begin{picture}(80,55)
\put(-1.5,3){\epsfig{file=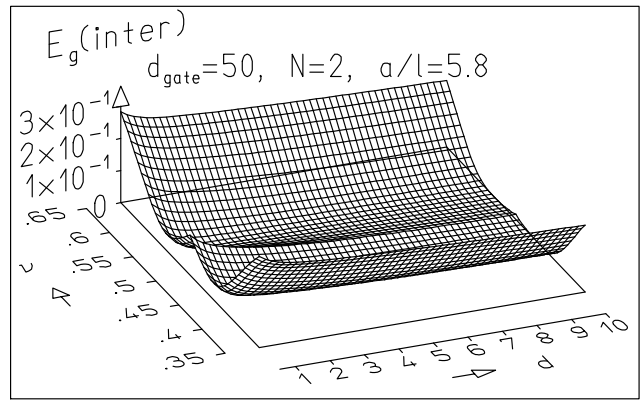,height=52mm}}
\end{picture}
\caption{Estimated charge gap due to interlayer back-scattering interactions.  These interactions
are always relevant and lead, in the absence of interlayer tunneling, to states with spontaneous
interlayer phase coherence.The energy scale in this figure is $\sim e^2/\epsilon \ell$ which
is $\sim k_B 50 {\rm K}$
for a typical higher Landau level experiment.  The energies should be reduced to account for
screening from inter-Landau level transitions that we have not included in our calculations.
}
\label{gap_inter}
\end{figure}

\begin{figure}
\unitlength=1mm
\begin{picture}(80,62)
\put(-1.5,3){\epsfig{file=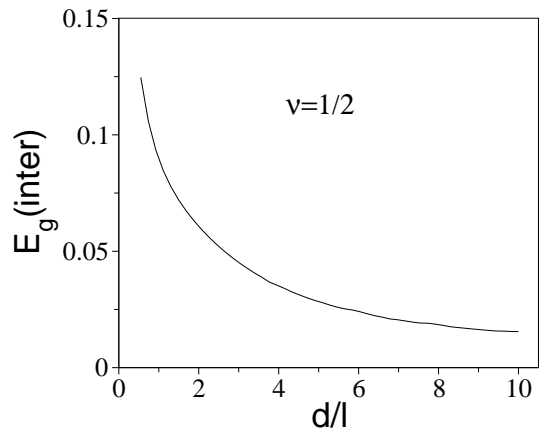,height=62mm}}
\end{picture}
\caption{Estimated charge gap due to interlayer back-scattering interactions, 
for balanced bilayers ($\nu=1/2$ in each layer), as a function of layer separation.
This dependence is extracted from Fig.~\ref{gap_inter} and shown here for clarity. 
}
\label{gap_inter2}
\end{figure}
For intralayer back scattering, the bare backscattering
interaction matrix element has both direct and
exchange contributions, while interlayer back-scattering has only a direct contribution.
An elementary calculation using Landau gauge basis states leads to the following explicit
expressions that were used to obtain gap estimates:
\\
\underline{Same-Layer Direct}
\bea
k_1 l^2 &=& Y_1 + Q l^2 \quad, \quad k_2 l^2 = Y_2 - Q l^2 \quad,
\\[3mm]
k_3 l^2 &=& Y_1 \quad , \quad k_4 l^2 = Y_2  \quad,
\eea

\begin{figure}
\unitlength=1mm
\begin{picture}(80,55)
\put(-1.5,3){\epsfig{file=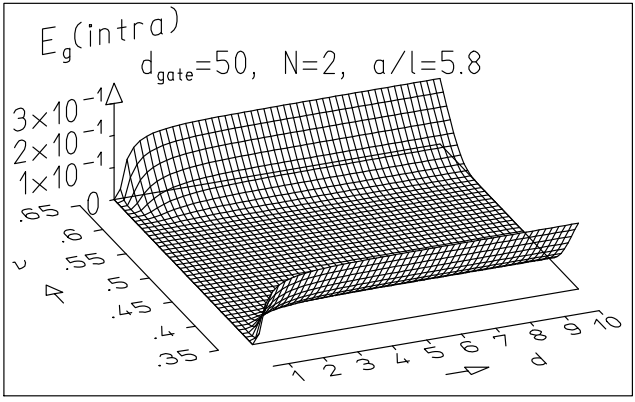,height=52mm}}
\end{picture}
\caption{Estimated charge gap that would result from intralayer back scattering interactions in
the bilayer case.  When the scaling dimension is larger than two, the gap vanishes.  Intralayer
interactions are more important than interlayer interactions only at very large layer separations.
The energy scale in this figure is $e^2/\epsilon \ell$.}
\label{gap_intra}
\end{figure}

and
\bea
&& \left< Y_1 + Ql^2,Y_2-Ql^2| V| Y_1,Y_2 \right> =
\frac{1}{2\pi} e^{-a^2 \nu^2/2l^2} \cdot
\nonumber\\[3mm]
&& \cdot\int \rd q_x  e^{-q_x^2 l^2/2} e^{-i q_x a} V_s^n(q_x,Q) \quad ,
\eea
\underline{Same-Layer Exchange}
\bea
k_2 l^2 &=& Y_1 + Q l^2 \quad, \quad k_1 l^2 = Y_2 - Q l^2 \quad,
\\[3mm]
k_3 l^2 &=& Y_1 \quad , \quad k_4 l^2 = Y_2  \quad,
\eea
\bea
&& \left< Y_2 - Ql^2,Y_1  + Ql^2| V| Y_1,Y_2 \right> =
\frac{1}{2\pi} e^{-a^2/2l^2} \cdot
\nonumber\\[3mm]
&& \cdot\int \rd q_x  e^{-q_x^2 l^2/2} e^{-i q_x a \nu} V_s^n(q_x,a/l^2) \quad ,
\eea
\underline{Different-Layer Direct}
\bea
k_1 l^2 &=& Y_1 + Q l^2 \quad, \quad k_2 l^2 = Y_1  \quad,
\\[3mm]
k_3 l^2 &=& Y_1 \quad , \quad k_4 l^2 = Y_1 + Q l^2 \quad,
\eea
\bea
&& \left< Y_2 ,Y_1| V| Y_1,Y_2 \right> =
\frac{1}{2\pi} e^{-a^2 \nu^2/2l^2} \cdot
\nonumber\\[3mm]
&& \cdot\int \rd q_x  e^{-q_x^2 l^2/2}  V_D^n(q_x,Q) \quad ,
\eea
where the subscripts $S$ and $D$ refer to two-dimensional Fourier transforms of the 
Coulomb interactions between electrons in same and different layers. 
We see in Fig.~\ref{gap_inter} and Fig.~\ref{gap_intra} that the importance of 
interlayer interactions diminishes rather slowly with layer separation, leading to 
sizable integer quantum Hall gaps out to large layer separations.

\begin{figure}
\unitlength=1mm
\begin{picture}(70,64)
\put(-3.5,6){\epsfig{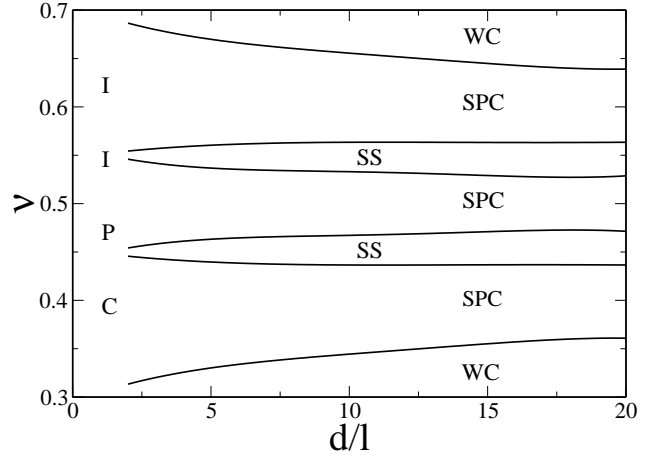}}
\end{picture}
\caption{
Apparent phase diagram predicted for experimental
studies of high mobility bilayer systems at dilution fridge temperatures. 
The various phases in this illustration have qualitatively different transport properties.
These calculations are for stripe states in the $N=2$ orbital Landau level ($a = 5.8 \ell$), 
with weak remote-gate screening ($d_{\rm gate} =50 \ell$). 
It is possible to explore the phase diagram experimentally in a single sample, since both the 
top layer filling factor $\nu$ and the normalized interlayer separation $d/\ell$, are 
altered when the charge imbalance and total electron density are changed by using 
front and back gates in combination.
For interlayer spacing $d$ less than approximately $1.5 \ell$ and any charge imbalance,
we expect the bilayer to be in an isotropic interlayer phase coherent (IIPC) state
which has a large gap, integer quantum Hall effect and isotropic transport properties. 
Anisotropic states are expected only for more widely spaced layers, $d> 1.5 \ell$.
For strongly unbalanced layers ($\nu$ far from 1/2) 
we expect anisotropic Wigner crystal (WC) states to appear because of intralayer backscattering
interactions, just as they do in the single layer case.  These states will 
exhibit a quantum Hall effect with an odd integer quantized Hall conductivity. 
Stripe (smectic metal) states (SS) tend to occur when each layer has a 
filling factor close to $\nu=1/2$, but as in the single layer case these
states are never the true ground states.  Smectic metal states show anisotropic transport but
do not show an integer quantum Hall effect.  Interlayer backscattering interactions
always induce charge gaps but these are sometimes too small to be observable at a typical dilution fridge 
temperatures which we take to be $0.001 e^2/\epsilon l$.   Regions with an estimated charge
gap larger than this value are labeled as smectic phase coherent (SPC) state regions in
the phase diagram.  Smectic phase coherent states have an odd integer quantum 
Hall effect, and are expected to have transport properties which are much more
anisotropic than those of the anisotropic Wigner crystal states.
This state should also exhibit giant interlayer tunneling conductance anomalies at low bias voltages.
}
\label{phase_diagram}
\end{figure}

Our results for the energy gaps are summarized in Fig.~\ref{phase_diagram}  by a schematic phase diagram
intended to represent predicted experimental findings in very high mobility bilayer 
systems at dilution refrigerator temperatures.  This phase diagram was constructed
from a recipe specified below.  Different regions of the phase diagram as a function of 
layer separation $d/l$ and imbalance, characterized by $\nu$, are identified as 
exhibiting the behavior of one of the following phases.  The bilayer smectic state 
is a state with no integer quantum Hall effect and anisotropic transport.  The 
coherent bilayer smectic state will have an integer quantum Hall effect but will
still have anisotropic transport at finite temperature.  The bilayer Wigner crystal state will
have an integer quantum Hall effect with an odd integer 
quantized Hall conductivity.   We predict bilayer smectic state behavior 
when neither interlayer nor intralayer back-scattering interactions produce a gap
larger than $0.001 e^2/\epsilon \ell$.  We judge that a gap smaller than this size
would not produce observable effects in a typical dilution fridge experiments. 
Interlayer backscattering interactions are much more effective than intralayer
interactions in producing gaps because they are strongly relevant.
We predict bilayer Wigner crystal behavior when the intralayer back-scattering yields the largest
gap and a gap that exceeds our minimum value.  These states are 
expected only when the charge imbalance is large or the layer separation is quite large.
We predict bilayer coherent smectic states when interlayer back-scattering produces
the largest gap, provided again that it exceeds our minimum value. 
Because the intralayer interactions are strongly relevant observable gaps are expected
out to very large layer separations, an unexpected result of our analysis.
The interval of charge imbalance where stripe (smectic metal) states are 
expected expands only modestly with layer separation, but {\em is} sensitive to the 
orbital index $N$ of the Landau levels, since nodes in the orbital wavefunctions can cause
the bare backscattering matrix element to vanish at particular $N$-dependent values of $\nu$.
The details of boundaries separating stripe state and stripe-phase-coherent regions of
this phase diagram will be quite different for different values of $N$. 
As we have emphasized, our approach is reliable only when quantum fluctuations around the 
mean-field stripe state of Hartree Fock theory are weak.  For small layer separations the 
charge gaps start to become comparable to the underlying microscopic energy scales.
In this regime we expect that the ground state is actually an isotropic coherent bilayer state,
but are unable to provide a reliable quantitative estimate of the layer separation at which
this transition occurs. 


\section{Discussion and Conclusions}

In this paper, we have studied double layer quantum Hall systems
at odd integer total filling fractions. 
Mean field theory predicts that these systems can form striped ground states.
This observation serves as the starting point for our work.  The Hilbert space 
in which the low-energy excited states of mean-field bilayer stripe states reside may be mapped
to those of an infinite set of coupled Luttinger liquids, one for each stripe,
allowing us to borrow bosonization techniques from the literature
on one-dimensional electron systems. Quantum fluctuations around the
mean-field stripe state are conveniently described
in terms of Bose quantum fields that can be interpreted as representing  
charge density and the position fluctuations along each stripe edge.
The interactions that control quantum
fluctuations in the electron ground state include both forward scattering terms
which contribute to quadratic interactions in the Boson Hamiltonian and 
weak, but more complicated back-scattering terms.
The coupled Luttinger liquid model obtained when the back-scattering 
interactions are neglected is not of the standard form because both charge and 
position terms in the effective Hamiltonian have a matrix character and because 
the energy cost of fluctuations in which stripes move collectively 
is small when the stripes are not pinned.  We find that the latter property
leads to Fermion spatial correlations whose decay is faster than any power law, to 
a specific heat that vanishes less quickly than $T$ for $T \to 0$, and to 
a tunneling density of states that vanishes faster than any power law for $E \to 0$. 
These properties of bilayer stripe states are similar to properties established
previously for single layers by Lopatnikova {\it et al.}
and Barci {\it et al.}.  There is no limit in which bilayer stripe quantum Hall states can 
be treated as a system of weakly coupled Luttinger liquids.

We address the role played by intralayer and interlayer back-scattering interactions
by evaluating their perturbative renormalization group scaling dimensions, following an 
approach two of us have taken previously for the case of single-layer stripe
states.\cite{MacDonald}  In the single-layer case we reached the conclusion that 
these interactions are always relevant, and that they likely drive the system to a 
Wigner crystal state with an energy gap.  Estimates of the size of this gap based on bare 
back-scattering matrix elements and scaling dimensions gave extremely small values, however, 
consistent with the observation of stripe state phenomenology at temperature scales that
could be reached experimentally.  Since other researchers have reached different conclusion
about the relevance of back-scattering interactions in single layer systems, it is worthwhile
in stating the conclusions we have reached in the present work to emphasize once again
the philosophy that underpins our calculations and explain why we have considerable
confidence in the conclusions we reached previously.

Our identification of a low-energy 
Hilbert space in which it is possible to derive a simplified many-electron Hamiltonian is
based on the experimental discovery of stripe states and on evidence from experiment
that the true ground state is energetically very close to the mean-field theory ground state.
In our view the most convincing evidence in this regard is the ability\cite{Jungwirthrecent}
of Hartree-Fock theory to accurately predict the dependence of the stripe state orientation
on in-plane field strength, quantum well width, and other microscopic parameters.  In single-layer
systems, quantum fluctuations are important only at low-energies and long length scales.
When mean-field theory accurately describes the microscopic length scale physics,
we can use the elementary excitations of the Hartree-Fock stripe state
to identify the Hilbert space of low-energy excitations, and confidently use
bare interaction matrix elements to estimate forward and back-scattering interaction 
parameters.  The issue of quantum stability of smectic states 
in single-layer systems has received interest 
partly because it is closely related to the possible existence\cite{Lubensky99,Fradkinold,Sondhi01}
of freely sliding analogs of the Kosterlitz-Thouless phase in stacked two-dimensional XY models.
Although it is certainly clear\cite{Fradkinold} that interacting Luttinger liquid fixed-point
actions exist for which back-scattering interactions are irrelevant, that is not sufficient to 
decide on their relevance in the case of quantum Hall stripe states.  Crudely speaking, 
irrelevance in the case of repulsive interactions requires\cite{MacDonald,singlelayer}
that the forward scattering interaction strength decay in a strongly
non-monotonic way with edge separation.  For single-layer systems Fertig and 
collaborators\cite{Yi00} have estimated forward scattering amplitudes using an approach that 
goes beyond the weak coupling approximations we employ, doing so however in a partially 
{\em ad hoc} manner by fitting their model to collective modes evaluated in a time-dependent
Hartree-Fock approximation.  Their conclusion on the relevance of back-scattering interactions 
is opposite to ours.  The source of the
discrepancy may be traced to broken particle-hole symmetry in the half-filled Landau level
Hartree-Fock approximation Wigner crystal state that they use to extract strong-coupling
interaction parameters.  {\em For a single layer stripe state, back-scattering 
interactions can be irrelevant only if the true ground state at $\nu=1/2$ breaks particle-hole
symmetry.}  This raises an interesting question.  Could there be another class of as yet 
undetected phase transitions that occur in the quantum Hall regime either in the high 
$N$ stripe state regime or for lower $N$? 
Broken particle-hole symmetry at $\nu=2$ would imply a finite-temperature 
phase transition in the 2D Ising universality class, for which the deviation of the 
Hall conductivity at $\nu=1/2$ from $e^2/2h$ could be taken as the order parameter.
There is certainly no evidence for such a phase transition in experiment, although it
might be washed out by disorder\cite{imrywortis} even if it occurred.  In any event, broken
particle-hole symmetry in the ground state would require that a phase transition occur 
between the high-temperature stripe state of Hartree-Fock theory that does not have broken
particle-hole symmetry and a low-temperature stripe state in which particle-hole symmetry
is broken and back-scattering is irrelevant.  In
light of the evidence that fluctuation corrections to the Hartree-Fock ground state are 
weak, we still believe that the simpler conclusion of our earlier work is more likely
to be correct, namely that particle-hole symmetry is not broken and that 
the smectic state is not stable.

As we have emphasized several times the approach we have taken does not lead to 
standard coupled Luttinger liquid behavior.  In particular, the decays of fermion 
correlation functions at large distances, and of the tunneling density of states at 
small energies, are faster than power laws.  This property occurs in our analysis because
of broken translational symmetry in the stripe state which makes its energy functional
invariant under a simultaneous translation of all stripes.  Barci {\it et al.}\cite{Barci},
have argued that this unusual property might signal a failure of the perturbative 
renormalization group transformation we have used, which rescales spatial coordinates
along the stripe edges but not across them.  Although
we agree that our conclusions concerning the nature of the true ground 
state could in principle be altered if it were possible to extend the perturbative RG analysis
to higher order.  Indeed this must happen when our analysis is applied to low index 
Landau levels in which stripe states do not occur.  We do not believe, however,
that the unusual correlation functions signal a greater
likelihood of this eventuality than normally applies to lowest order perturbative 
RG calculations. In practice, the microscopic back-scattering amplitudes treated perturbatively 
are sufficiently weak in high Landau levels that our lowest order calculations seem likely to describe
what happens down to the lowest temperatures available experimentally.  In our view
the approach we have taken should be trusted when experimental evidence suggests that 
the physics at low-energies is described by the stripe states of Hartree-Fock theory.

We find here that the role of back-scattering interactions is quite different in the 
bilayer case compared to the single layer case.  At very large layer separations,
the single-layer case in which stripe state physics occurs down to very low temperatures
for $\nu \sim 1/2$ is recovered.  However, already for layer separations $\sim 10 \ell$,
we find that interlayer back-scattering interactions which drive the system toward a 
state with spontaneous interlayer phase coherence along the edges become important and lead to a state 
with a substantial charge gap.  Our prediction of odd integer quantum Hall effects with 
anisotropic finite-temperature transport coefficients in surprisingly widely separated 
bilayer systems is an important result of this paper.
This conclusion about the properties of spontaneously coherent
stripe states in the absence of interlayer tunneling differs from that reached by Fertig and 
collaborators who, incorrectly in our view, ignore inter-edge coupling in considering the 
properties of coherent stripes.  Interestingly, intra-layer back-scattering interactions that 
drive the system toward a Wigner crystal state are irrelevant in this regime.  We conclude
that stripe states are indeed stable in bilayer quantum Hall systems, unlike
the single layer case, but not smectic metal states.
It seems likely that for very small layer separations, back-scattering
interactions will drive the system toward a uniform charge density state with interlayer coherence,
although our perturbative approach is not able to offer any substantial guidance in deciding
this question.   

The study of stripe state physics in single-layer quantum Hall systems requires samples
of exceptional quality, beyond that required for studies of fractional quantum Hall physics
with lower index partially filled Landau levels which can be studied at higher magnetic
fields.  It is still not
possible to create bilayer quantum Hall systems with disorder that is as weak as that in 
single layer quantum Hall systems.  Nevertheless recent samples appear to 
be of a quality that opens the physics of stripe states in bilayer systems up to experimental
study.    We expect on the basis of this work, and of previous theoretical work, that
the physics will be rich, with much potential for surprises beyond the properties anticipated here. 

The authors are grateful for helpful and stimulating interactions with 
Dave Allen, Alan Dorsey, Rene C\^ ot\' e,
Herb Fertig, Eduardo Fradkin, Steve Kivelson, 
Tom Lubensky, Tilo Stroh, Alexei Tsvelik,  
and Carlos Wexler.  Work in Austin was supported by the National Science 
Foundation under grant NSF-DMR-0115947.  Work in Santa Barbara 
was supported by NSF grants DMR-0210790 and Phy-9907949.


\end{document}